\documentclass[letterpaper, twocolumn, 10pt]{article}

\usepackage{tikz}
\usepackage{amsmath}
\usepackage{listings}
\usepackage{array}
\usepackage{multirow}
\usepackage{enumitem}
\usepackage{pifont}
\usepackage[breakable]{tcolorbox}
\usepackage{adjustbox}
\usepackage{booktabs}
\usepackage{algorithm}
\usepackage{makecell}
\usepackage[noend]{algpseudocode}
\usepackage{usenix}
\usepackage{xspace}
\usepackage{url}
\usepackage{caption}
\usepackage{amsfonts}
\usepackage{acronym}
\usepackage{colortbl}
\definecolor{mygreen}{RGB}{255, 218, 185}
\definecolor{myblue}{RGB}{202, 220, 255}

\definecolor{verylightgray}{rgb}{.97,.97,.97}
\lstdefinelanguage{Solidity}{ keywords=[1]{anonymous, assembly, assert, balance, break, call, callcode, case, catch, class, constant, continue, constructor, debugger, default, delegatecall, delete, do, else, emit, event, experimental, export, external, false, finally, for, function, gas, if, implements, import, in, indexed, instanceof, interface, internal, is, length, library, log0, log1, log2, log3, log4, memory, modifier, new, payable, pragma, private, protected, public, pure, push, require, return, returns, revert, selfdestruct, send, solidity, storage, struct, suicide, super, switch, then, this, throw, transfer, true, try, typeof, using, value, view, while, with, addmod, ecrecover, keccak256, mulmod, ripemd160, sha256, sha3}, 
keywordstyle=[1]
\color{blue}
\bfseries, keywords=[2]{address, bool, byte, bytes, bytes1, bytes2, bytes3, bytes4, bytes5, bytes6, bytes7, bytes8, bytes9, bytes10, bytes11, bytes12, bytes13, bytes14, bytes15, bytes16, bytes17, bytes18, bytes19, bytes20, bytes21, bytes22, bytes23, bytes24, bytes25, bytes26, bytes27, bytes28, bytes29, bytes30, bytes31, bytes32, enum, int, int8, int16, int24, int32, int40, int48, int56, int64, int72, int80, int88, int96, int104, int112, int120, int128, int136, int144, int152, int160, int168, int176, int184, int192, int200, int208, int216, int224, int232, int240, int248, int256, mapping, string, uint, uint8, uint16, uint24, uint32, uint40, uint48, uint56, uint64, uint72, uint80, uint88, uint96, uint104, uint112, uint120, uint128, uint136, uint144, uint152, uint160, uint168, uint176, uint184, uint192, uint200, uint208, uint216, uint224, uint232, uint240, uint248, uint256, var, void, ether, finney, szabo, wei, days, hours, minutes, seconds, weeks, years}, 
keywordstyle=[2]
\color{teal}
\bfseries, keywords=[3]{block, blockhash, coinbase, difficulty, gaslimit, number, timestamp, msg, gas, sender, sig, value, now, tx, gasprice, origin}, 
keywordstyle=[3]
\color{violet}
\bfseries, keywords=[4]{mload,create,create2,add,contract,receive，sub}, keywordstyle=[4]
\color{red}
\bfseries, identifierstyle=
\color{black}
, sensitive=true, comment=[l]{//}, morecomment=[s]{/*}{*/}, commentstyle=
\color{gray}
\ttfamily, stringstyle=
\color{red}
\ttfamily, morestring=[b]', morestring=[b]", }

\lstdefinestyle{codestyle}{ language=Solidity, basicstyle=\ttfamily\scriptsize, keywordstyle=
\color{blue}
, commentstyle=
\color{green!60!black}
, stringstyle=
\color{red}
, breaklines=true, numbers=left, numberstyle=\tiny
\color{gray}
, stepnumber=1, frame=single, xleftmargin=10pt,
captionpos=b, }
\setlist[itemize]{noitemsep,nolistsep,left=0pt}
\setlist[enumerate]{noitemsep,nolistsep,left=0pt}
\usepackage{titlesec}
\titlespacing*{\paragraph}{0pt}{0ex}{1em}

\usepackage{caption}
\acrodef{LLM}{Large Language Model}
\acrodef{SAST}{Static Application Security Testing}
\acrodef{DAST}{Dynamic Application Security Testing}
\acrodef{PoC}{Proof-of-Concept}
\acrodef{SOTA}{State-of-the-Art}

\begin{document}

	\date{}

	\newcommand{\tool}{\textsc{A2}\xspace}

	\title{Agentic Discovery and Validation of Android App Vulnerabilities}
	\author{
    {\rm Ziyue Wang}\\
	Nanjing University
	\and
	{\rm Liyi Zhou}\\
	The University of Sydney, Decentralized Intelligence AG, UC Berkeley RDI
	} 

	\maketitle

	\begin{abstract}
		Existing Android vulnerability detection tools overwhelm teams with thousands of low-signal warnings
		yet uncover few true positives. Analysts spend days triaging these results, creating a bottleneck
		in the security pipeline. Meanwhile, genuinely exploitable vulnerabilities often slip through,
		leaving opportunities open to malicious counterparts.

		We introduce \tool{}, a system that mirrors how security experts analyze and validate
		Android vulnerabilities through two complementary phases: \emph{(i)} \textit{\underline{A}gentic
		Vulnerability Discovery}, which reasons about application security by combining semantic
		understanding with traditional security tools; and \emph{(ii)} \textit{\underline{A}gentic
		Vulnerability Validation}, which systematically validates vulnerabilities across Android's
		multi-modal attack surface—UI interactions, inter-component communication, file system
		operations, and cryptographic computations.

		On the Ghera benchmark ($n{=}60$), \tool{} achieves $78.3\%$ coverage, surpassing state-of-the-art
		analyzers (e.g., APKHunt $30.0\%$). Rather than overwhelming analysts with thousands of warnings,
		\tool{} distills results into 82 speculative vulnerability findings, including~$47$ Ghera
		cases and~$28$ additional true positives. Crucially, \tool{} then generates working \acp{PoC}
		for $51$ of these speculative findings, transforming them into validated vulnerability findings
		that provide direct, self-confirming evidence of exploitability.


		In real-world evaluation on 169 production APKs, \tool{} uncovers 104 true-positive zero-day
		vulnerabilities. Among these, 57 ($54.8\%$) are self-validated with automatically generated \acp{PoC},
		including a medium-severity vulnerability in a widely used application with over 10 million installs.
	\end{abstract}

	\section{Introduction}

	Mobile applications are the primary interface to many digital services. The Google Play Store hosts
	approximately~$2.8$ million Android applications~\cite{GooglePlayConsole2024}. These
	applications process sensitive user and enterprise data and implement business-critical
	workflows, making them attractive targets for adversaries. Reported mobile security incidents
	increased by~$32\%$ year over year~\cite{verizon2024msi}. The scale of Android applications,
	together with their privileged access to device resources and user data, complicate comprehensive
	security assessment.

\begin{figure}[t]
		\centering
		\includegraphics[width=\columnwidth]{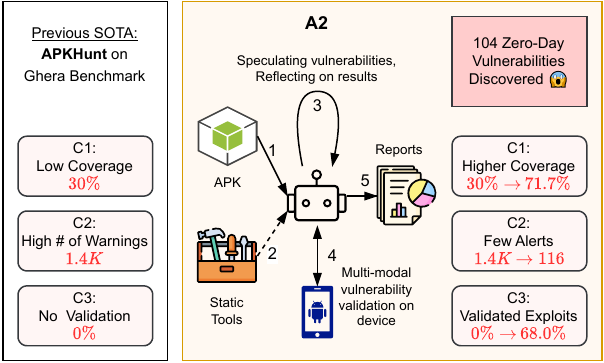}
		\caption{Android vulnerability assessment faces three key challenges: limited coverage,
		excessive warnings, and lack of validation. \tool{} tackles these by mimicking expert analysis
		through \textit{Agentic Vulnerability Discovery} and \textit{Agentic Vulnerability
		Validation} in four steps: (1) APK input, (2, optional) integration of static tool signals, (3)
		speculation on candidate vulnerabilities with multi-modal validation, and (4) generation of
		final reports. Compared to prior SOTA (APKHunt), \tool{} increases coverage ($3 0\% \rightarrow
		71.7\%$, Table~\ref{tab:vuln_detect_summary} in the Appendix), while reducing from 1.4K
		warnings to 116 vulnerability findings, and enabling exploit validation ($0\% \rightarrow 68.
		0\%$, Table~\ref{tab:poc_summary} in the Appendix). Beyond the Ghera benchmark, applying
		\tool{} to real-world apps revealed 104 previously unknown zero-day vulnerabilities (Section~\ref{sec:real_world_vuln_detection}).}
		\label{fig:motivation}
	\end{figure}

	Prior work on APK vulnerability detection has largely relied on \ac{SAST} tools~\cite{owasp_sast_definition},
	such as FlowDroid~\cite{flowdroid}, MobSF~\cite{mobsf}, and APKHunt~\cite{apkhunt}. While useful
	for surfacing known patterns, these tools leave material gaps in end-to-end security validation (cf.
	Figure~\ref{fig:motivation}).

	\begin{itemize}[noitemsep,nolistsep]
		\item \textbf{C1. Low Coverage}: \ac{SAST} tools match predefined patterns and data-flow
			rules. However, many Android vulnerabilities are context and execution dependent, and thus
			may evade static, rule-based detection systems.

		\item \textbf{C2. High Volume of Warnings}: \ac{SAST} scans entire APKs, including dependencies,
			third-party libraries, and low-severity syntax issues, generating large volumes of low-signal
			warnings that overwhelm analysts and obscure issues in application code, driving costly manual
			triage~\cite{votipka2020understanding,Johnson:2013:WDS}.

		\item \textbf{C3. No Exploit Validation}: \ac{SAST} flags potential issues without writing
			\ac{PoC} code to demonstrate and validate exploitability. Meanwhile, \ac{DAST} tools~\cite{owasp_dast_definition}
			in theory can generate \acp{PoC}, but they struggle with Android's structured, multimodal
			input space (UI events, Intents\footnote{In Android, an \emph{Intent} is a message
			object that allows app components to request actions from each other. For example, an
			app can issue an implicit intent such as \texttt{ACTION\_VIEW} with a URL to ask the system
			to open the link in a browser. While convenient, intents can also introduce security risks.},
			filesystem, databases, shell), yielding a combinatorial search space that is challenging
			for fuzzers to explore effectively~\cite{10504267}.
	\end{itemize}

	Recent progress in \acp{LLM} marks a fundamental shift in Android security analysis, overcoming long-standing
	limitations of traditional techniques. We present \tool{}, a framework that mirrors the
	reasoning process of human analysts. At its core, \tool{} operates through two complementary phases:
	\textit{Agentic Vulnerability Discovery} and \textit{Agentic Vulnerability Validation}. The discovery
	phase reasons about application security, combining semantic code understanding with traditional
	security tools to form actionable vulnerability hypotheses. Building on these insights, the
	assessment phase systematically validates each hypothesis through coordinated agents that plan, execute,
	and verify exploitation attempts. Conceptually, \tool{} mirrors how security experts work: first
	understanding the application's security posture comprehensively, then methodically proving or disproving
	each potential vulnerability through hands-on testing. The result is a system that not only
	identifies vulnerabilities but demonstrates their real-world impact through concrete, verifiable
	attack paths. Our primary contributions are:

	\begin{itemize}
		\item \textbf{Human-Level End-to-end Security Analysis}: To our knowledge, \tool{} is the first
			system to mimic expert practices in vulnerability assessment. It performs \textit{Agentic
			Vulnerability Discovery} by combining semantic reasoning with traditional signals, and
			\textit{Agentic Vulnerability Validation} by planning, executing, and verifying exploits.
			This human-like approach enables adaptive discovery beyond static rules while ensuring
			rigorous validation with concrete evidence.

		\item \textbf{New \ac{SOTA} on Ghera Benchmark}: On 60 vulnerable APKs, we evaluate \tool{} alongside
			four reasoning \acp{LLM} and three \ac{SAST} baselines. Among single models, o3 attains 71.7\%
			(43/60). The ensemble of Gemini 2.5 Flash, Gemini 2.5 Pro, and o3 reaches 78.3\% (47/60)
			and covers all $27$ vulnerabilities detected by the \ac{SAST} tools. \tool{} aggregates $8
			,528$ outputs into $82$ speculative vulnerability findings while retaining all true positives.
			This shows that combining reasoning models with aggregation yields both higher coverage
			and significantly fewer speculative findings than existing approaches.

		\item \textbf{LLM-based Vulnerability Validation}: We evaluate \tool{} by generating \ac{PoC}
			exploits for~$82$ findings \tool{} reports on the Ghera benchmark. Within 20 iterations,
			using Gemini~2.5 Pro as the \textit{PoC Planner} and Gemini~2.5 Flash as the \textit{Task
			Executor} and \textit{Validator}, \tool{} achieves a 61.3\% success rate (46/75) on
			actionable vulnerabilities and correctly rules out all false positives (7/7). The \textit{Executor}
			averages~$4.49$ function calls per task, while the \textit{Validator} filters 12.6\% of erroneous
			claims through dynamically generated oracles, mitigating hallucination-induced false
			positives. When both the \textit{Executor} and \textit{Validator} are upgraded to Gemini~2.5
			Pro, the success rate increases to 68.0\% (+6.7\%), and erroneous claims drop to 4.7\% (-7.9\%).
			Overall, these results demonstrate a shift from simple detection toward end-to-end validation
			of vulnerabilities through executable evidence.

		\item \textbf{Zero Days}: We deploy \tool{} on 169 AndroZoo~\cite{allix2016androzoo} APKs
			released in 2024–2025, to assess \tool's effectiveness in practice. The vulnerability
			detection module reports 136 vulnerabilities, from which the agentic vulnerability validation
			module invalidates 29 false positives. Among the remaining 107 speculative vulnerability
			findings (104 true positives), \tool{} achieves a 54.8\% success rate (57/104) in end-to-end
			exploitation using Gemini 2.5 Pro as the \textit{PoC Planner} and Gemini 2.5 Flash as
			both \textit{Task Executor} and \textit{Task Validator}. We responsibly disclosed these zero-days,
			including cases in apps with over 10 million downloads.
	\end{itemize}

	\section{Background}
	This section provides background on Android security.

	\paragraph{Android Component Architecture}
	Android applications utilize a component-based architecture with four building blocks:
	Activities (UI screens), Services (background tasks), Broadcast Receivers (system events), and Content
	Providers (data access)~\cite{android_components}. These components communicate through Intents,
	which can be either explicit (targeting specific components) or implicit (system-resolved based
	on declared capabilities). The Android manifest file serves as the central configuration
	document, declaring all components along with their required permissions and export policies,
	effectively defining the application's security boundaries. This modular design enables component
	reuse across applications, but simultaneously introduces security risks. Improperly configured exported
	components can be accessed by malicious applications, while malformed intents can lead to
	attacks such as intent spoofing and privilege escalation~\cite{dvaSecurity2024}.

	\paragraph{Android Application Vulnerability Landscape}
	The Android ecosystem faces a diverse landscape of security vulnerabilities that compromise user
	privacy and system integrity. A comprehensive empirical study by VulsTotal~\cite{10.1109/TSE.2024.3488041}
	systematically categorizes Android application vulnerabilities into 58 distinct types across six
	major categories: cryptographic weaknesses, inter-component communication flaws, networking vulnerabilities,
	permission misconfigurations, storage issues, and web-related problems. This taxonomy reveals the
	complexity of Android security, where vulnerabilities often arise from the interaction between
	multiple system components rather than isolated code defects. Traditional detection approaches focus
	primarily on well-known vulnerability patterns, but complex security flaws require deep semantic
	understanding of application logic and cross-component data flows that challenge conventional
	automated analysis techniques.

	\paragraph{Static Application Security Testing Tools}

	\ac{SAST} tools~\cite{owasp_sast_definition} are designed to statically analyze source code or compiled
	versions of code to identify security flaws. According to OWASP~\cite{owasp_static_code_analysis},
	static code analysis is performed as part of white-box testing during the implementation phase,
	using techniques such as taint analysis and data flow analysis to highlight possible
	vulnerabilities within non-running source code. These tools form the backbone of modern Android application
	vulnerability detection, providing automated analysis capabilities across diverse application
	codebases. Popular Android \ac{SAST} tools include both shallow analysis tools such as MobSF~\cite{mobsf},
	APKHunt~\cite{apkhunt}, QARK~\cite{qark}, and AndroBugs~\cite{androbugs}, and deep taint
	analysis tools such as FlowDroid~\cite{flowdroid}, Amandroid~\cite{amandroid}, and DroidSafe~\cite{droidsafe}.

	\paragraph{LLM-Powered Agents in Security}

	\acp{LLM} have shown strong code comprehension and analysis skills, leading to their growing use
	in software security~\cite{llm4vulnerability2024}. Moving beyond rule-based tools, \ac{LLM}-powered
	agents leverage contextual reasoning to perform multi-step security analysis with minimal supervision.
	Unlike static analysis tools that depend on fixed vulnerability patterns, they adapt to new vulnerability
	types by understanding semantic code relationships. Advances in prompt engineering, tool integration,
	and multi-agent systems have enabled tasks such as vulnerability classification, exploit generation,
	and automated penetration testing~\cite{gptScan2023,AutoDroid}. Yet, \ac{LLM} agents face limits
	in context windows, hallucination risks, and variable quality. Research addresses these through
	verification mechanisms and hybrid designs that combine \ac{LLM} reasoning with traditional analysis~\cite{autonomous_exploit2024}.

	\paragraph{Multi-Agent Orchestration}
	LangGraph~\cite{langgraph} is a framework for organizing multi-agent workflows using graph
	structures. It represents agents as nodes and their interactions as edges, allowing developers
	to define how information and control flow between components. In practice, LangGraph acts as a
	programmable state machine: it coordinates message passing, supports looping and branching, and manages
	task execution order. While it is often presented as a new paradigm, its foundations align with long-standing
	concepts of workflow orchestration and state transition systems. Its main contribution is
	providing a convenient abstraction layer that makes it easier to connect \acp{LLM} with external
	tools, to structure multi-step reasoning pipelines, and to implement such systems in practice.
	In this work, \tool adopts LangGraph as it reduces our development overhead.

	\begin{figure*}[t]
		\centering
		\includegraphics[width=\linewidth]{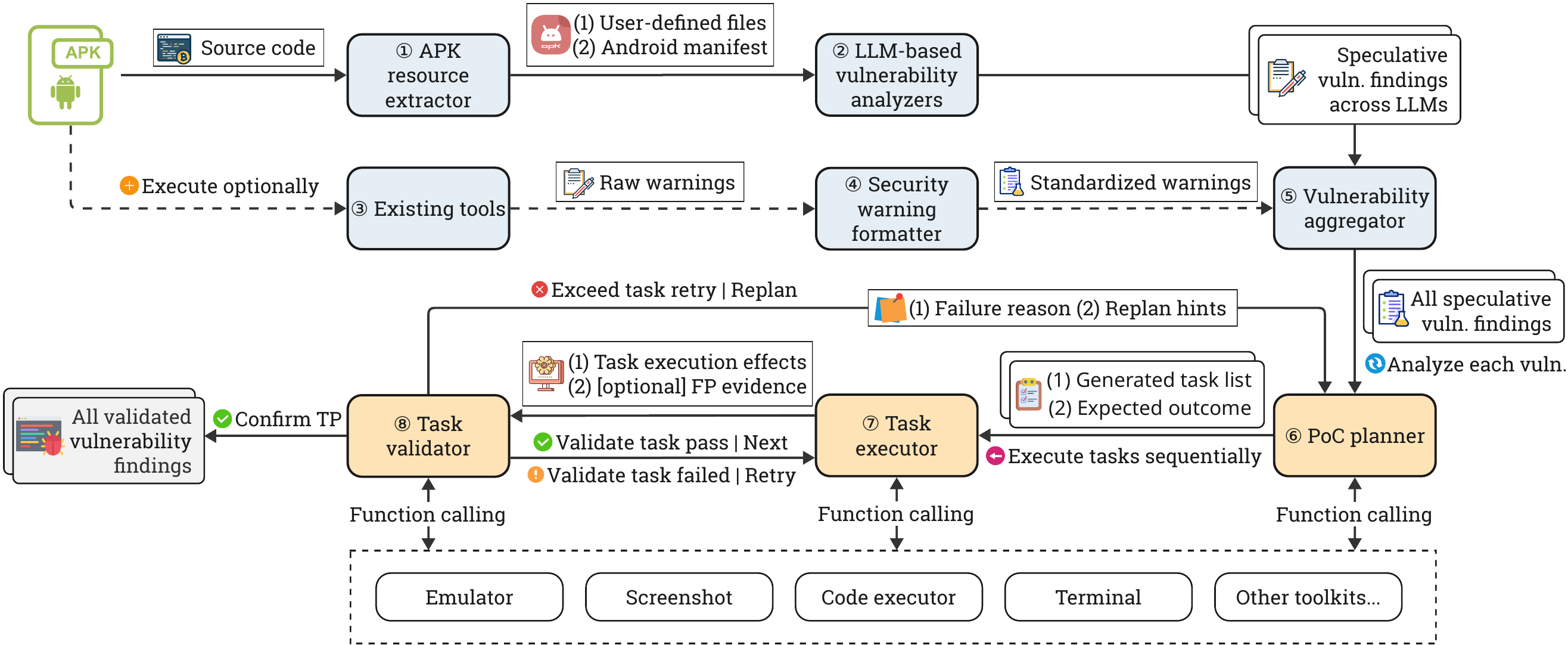}
		\caption{The end-to-end workflow of \tool, from APK input to validated vulnerability findings
		with proof-of-concept output. \tool operates in two phases: \textbf{Agentic Vulnerability
		Discovery} (\ding{192}-\ding{196}, \colorbox[HTML]{E6EDF5}{blue blocks}) and \textbf{Agentic
		Vulnerability Validation} (\ding{197}–\ding{199}, \colorbox[HTML]{F9E4BC}{brown blocks}). In
		the vulnerability discovery phase, the target APK is decompiled and its resources extracted (\ding{192}).
		LLM agents then analyze the application (\ding{193}), producing speculative vulnerability findings.
		Optionally, warnings from existing \ac{SAST} tools (\ding{194}) are passed through the formatter
		(\ding{195}), which standardizes them and produces additional speculative vulnerability findings
		to enhance coverage and confidence. All speculative findings are then consolidated by the
		aggregator (\ding{196}). In the vulnerability validation phase, each speculative vulnerability
		finding is passed to the PoC planner (\ding{197}), which generates task lists with expected outcomes.
		The task executor (\ding{198}) sequentially carries out these tasks through function calls (e.g.,
		emulator, screenshot analysis, code execution, terminal commands, and etc.) Finally, the task
		validator (\ding{199}) independently verifies outcomes using multiple oracles, providing feedback
		for iterative refinement until either successful validation or retry limits are reached.}
		\label{fig:architecture}
	\end{figure*}

	\section{Models}
	\label{sec:threat-model} This section defines the models and analysis scope.

	\begin{table}[t]
	\centering
	\caption{Examples of tools that both \tool{} and adversaries may use.}
	\label{tab:exploiter-toolkit}
	\footnotesize
	\begin{tabular}{ll}
		\toprule 
		\textbf{Capability Category} & \textbf{Example Supporting Tools} \\
		\midrule 
		Code Analysis & Jadx~\cite{jadx} \\
		Manifest Inspection & Androidguard~\cite{androidguard} \\
		Static Vulnerability Scanning & MobSF~\cite{mobsf}, APKHunt~\cite{apkhunt} \\
		Dynamic Runtime Control & Android Emulator~\cite{android_emulator} \\
		Large Language Models & OpenAI, Google Gemini, etc. \\
		\bottomrule
	\end{tabular}
\end{table}

	\paragraph{System Model}

	We model the Android ecosystem through three actors: \textit{(i)} developers, \textit{(ii)} distribution
	channels, and \textit{(iii)} end-user devices. Developers build applications that may contain
	vulnerabilities due to coding mistakes, misuse of APIs, or insecure third-party libraries. Applications
	are packaged as signed APKs, with identity tied to the signing key across updates. Source code is
	often closed, limiting visibility into developer practices. APKs are distributed through Google Play,
	third-party markets, and sideloading. For this study, we use AndroZoo~\cite{allix2016androzoo},
	a growing dataset aggregating APKs from major markets, offering a representative sample of real-world
	apps. On devices, we assume each app executes in a sandbox defined by UID and SELinux, with
	permissions mediating access to system resources and co-resident apps. Apps interact with local components,
	embedded web content, and remote backend services, which define the trust boundaries in our
	model: per-app sandboxing, inter-app communication, web content, and network channels. We also assume
	stock, non-rooted Android devices with the platform security model enforced, and exclude physical
	access and custom ROMs.

	\paragraph{Threat Model}

	We consider adversaries who attempt to exploit Android applications to steal sensitive data,
	escalate privileges across application boundaries, or tamper with application logic and communication
	channels. Such adversaries may start with only the APK, without access to source code or developer
	keys, but can analyze the package, execute it on a device or emulator, and craft exploit
	sequences that cross trust boundaries. We assume adversaries are capable of reverse engineering APKs,
	observing runtime behavior, and injecting crafted inputs through Android's interaction channels,
	including Intents, UI events, filesystem operations, and network traffic. They do not control
	the Android platform, kernel, or hardware. Attacks requiring rooted devices, custom firmware, or
	hardware side channels are out of scope. Adversaries instead focus on application-layer
	vulnerabilities introduced by developers or insecure library use. Functional bugs~\cite{demissie2025vlmfuzz,liu2024visiondroid,xiong2023empirical}
	that do not compromise security properties are considered out of scope. To support these
	capabilities, adversaries may rely on existing tools (cf. Table~\ref{tab:exploiter-toolkit}), such
	as decompilers, manifest inspection utilities, static analysis suites, and emulators, which also
	form the basis for our AI agent (\tool{}) when reasoning about attack strategies.

	\paragraph{Terminology}
	To ensure precise language throughout this paper, we distinguish between three related terms:

	\begin{enumerate}
		\item \textbf{Warning.} A potential issue reported by static analysis or rule-based tools. Warnings
			may include false positives and are not supported by a \ac{PoC} implementation. They require
			manual inspection to determine their relevance.

		\item \textbf{Speculative Vulnerability Finding.} A candidate vulnerability generated by
			\tool{}. Unlike warnings, these findings include a reasoned hypothesis about why the issue
			may be exploitable, but they remain unconfirmed until validation.

		\item \textbf{Validated Vulnerability Finding.} A vulnerability for which at least one
			working \ac{PoC} has been produced, and an \ac{LLM} assessment indicates that the \ac{PoC}
			demonstrates the vulnerability. While this stage goes beyond hypothesis, such findings
			may still include false positives if the \ac{PoC} or validation is misleading or incomplete.
	\end{enumerate}

	\section{\tool Design}

	Figure~\ref{fig:architecture} illustrates the end-to-end workflow of \tool. Given an APK as input,
	\tool performs code analysis to generate speculative vulnerability findings. Through iterative \ac{PoC}
	refinement and execution, \tool then transforms speculative vulnerability findings into validated
	vulnerability findings.

	\subsection{Agentic Vulnerability Discovery}
	\label{sec:hybrid_detection}

	\tool discovers vulnerabilities primarily through \acp{LLM}, with optional support from static
	tools and vulnerability databases.

	\subsubsection{\ac{LLM}-based Vulnerability Analyzer}
	Our analyzer consists of three components (cf. Figure~\ref{fig:architecture}):

	\paragraph{\ding{192} APK Resource Extractor}
	\textit{(i)} employs Jadx~\cite{jadx} to decompile bytecode into source code, then \textit{(ii)}
	excludes third-party libraries and Android framework code, and \textit{(iii)} extracts manifest details
	(e.g., component declarations, permission specifications, and inter-component communication
	policies.)

	\paragraph{\ding{193} LLM Vulnerability Analyzer}
	processes the code and manifest data using \acp{LLM} to identify security flaws and vulnerability
	patterns that may evade traditional rule-based detection. It generates standardized reports with
	exact line-level mappings in code, enabling the discovery of previously unseen vulnerabilities and
	context-dependent flaws that require understanding of application logic and data flow.

	\subsubsection{(Optional) Third-party Tool Integration}

	\tool can integrate third-party tools that accept an APK as input.

	\paragraph{\ding{194} Existing Static Tools}
	such as MobSF~\cite{mobsf} and APKHunt~\cite{apkhunt} offer two advantages over pure \acp{LLM}: \textit{(i)}
	they provide precise code locations and vulnerability context that enrich the evidence base, and
	\textit{(ii)} their deterministic data flow analysis guides targeted exploitation by revealing
	concrete attack paths. This integration complements the semantic reasoning of \acp{LLM} with the
	precision of static analysis, enabling broader and more reliable vulnerability discovery.

	\paragraph{\ding{195} Security Warning Formatter}
	standardizes the diverse outputs of static analysis tools, which often vary in format and terminology.
	Leveraging \acp{LLM}, it harmonizes formats, aligns semantics, and structures reports into
	consistent security warnings for downstream processing.

	\subsubsection{Vulnerability Aggregation}

	The discovery process concludes with \ding{196}, the \textbf{Vulnerability Aggregator}, an \ac{LLM}-based
	component that unifies outputs from \ding{193} and \ding{195}. The aggregator performs three key
	functions: \textit{(i)} \textbf{Security issue filtering}, removing warnings unrelated to exploitable
	flaws such as code quality or performance issues; \textit{(ii)} \textbf{Semantic deduplication},
	collapsing duplicate findings across tools through similarity analysis and contextual reasoning;
	and \textit{(iii)} \textbf{Evidence synthesis}, merging complementary signals to construct richer
	vulnerability profiles. Together, these functions yield \textit{speculative vulnerability
	findings}, each representing a reasoned hypothesis about potential exploitable flaws, ready for
	validation in Section~\ref{sec:agentic_vulnerability_validation}.

	\subsection{Agentic Vulnerability Validation}
	\label{sec:agentic_vulnerability_validation}

	The Agentic Vulnerability Validation phase employs a multi-agent system to confirm speculative findings
	through \ac{PoC} exploits. \tool{} uses LangGraph~\cite{langgraph} for stateful coordination and
	task routing (Table~\ref{tab:comprehensive_toolkit}). At its core, \tool runs a feedback loop: the
	\textit{\ac{PoC} Planner} breaks down a strategy into verifiable tasks, the \textit{Task
	Executor} carries them out, and the \textit{Task Validator} independently judges the results as
	success or failure.

	\begin{table}[t]
	\centering
	\caption{Available function calls for Agentic Vulnerability Validation. We provide~$29$ specialized
	functions grouped into eight categories: \textit{(i)} code \textbf{execution} (dynamic Python scripts,
	package management), \textit{(ii)} device \textbf{control} (hardware key simulation, ADB shell
	commands), \textit{(iii)} \textbf{file} system operations (local and device-side file management
	for extraction and analysis), \textit{(iv)} \textbf{code} navigation (source inspection),
	\textit{(v)} \textbf{UI} interaction, \textit{(vi)} \textbf{log} analysis (logcat filtering),
	\textit{(vii)} \textbf{APK} generation for custom test payloads, and \textit{(viii)} \textbf{web}
	server management for simulating external services or adversarial endpoints. The function calls
	design follows three principles: functional decomposition (clear domains and separation of
	concerns), role-based access control (agents restricted to curated tool subsets), and isolation/sandboxing
	(dangerous actions run in safe environments). A modular structure allows extension as new
	validation techniques emerge. Functions marked as \textit{read-only} -- highlighted
	in {\color{blue}blue} -- do not modify the runtime state of the target application inside the Android
	emulator, though they may still affect the local analysis environment}.
	\label{tab:comprehensive_toolkit} \resizebox{\columnwidth}{!}{
	\begin{tabular}{lllccc}
		\toprule                                   & \textbf{Tool Function}                  & \textbf{Description}                            & \rotatebox{90}{\textbf{Planner}} & \rotatebox{90}{\textbf{Executor}} & \rotatebox{90}{\textbf{Validator}} \\
		\midrule \multirow{2}{*}{\textbf{Execute}} & {\color{blue}execute\_python\_script}   & Execute Python code in VM                       &                                  & \checkmark                        & \checkmark                         \\
		                                           & {\color{blue}install\_python\_package}  & Install Python packages in VM                   &                                  & \checkmark                        & \checkmark                         \\
		\midrule \multirow{4}{*}{\textbf{Control}} & press\_hardware\_key                    & Press system keys (e.g., back, home, etc.)      &                                  & \checkmark                        &                                    \\
		                                           & {\color{blue}launch\_application}       & Start application with optional activity        & \checkmark                       & \checkmark                        &                                    \\
		                                           & restart\_application                    & Stop and restart application with activity      &                                  & \checkmark                        &                                    \\
		                                           & execute\_shell\_command                 & Execute ADB shell commands                      &                                  & \checkmark                        & \checkmark                         \\
		\midrule \multirow{7}{*}{\textbf{File}}    & {\color{blue}pull\_device\_file}        & Transfer file from Android device to host       & \checkmark                       & \checkmark                        & \checkmark                         \\
		                                           & upload\_file\_to\_device                & Upload file from host to Android device         &                                  & \checkmark                        &                                    \\
		                                           & {\color{blue}check\_file\_existence}    & Verify file or directory existence on device    & \checkmark                       & \checkmark                        & \checkmark                         \\
		                                           & {\color{blue}analyze\_file\_content}    & Search file content using regex patterns        & \checkmark                       & \checkmark                        & \checkmark                         \\
		                                           & {\color{blue}create\_local\_file}       & Create file in execution environment            &                                  & \checkmark                        & \checkmark                         \\
		                                           & {\color{blue}read\_local\_file}         & Read file from execution environment            & \checkmark                       & \checkmark                        & \checkmark                         \\
		                                           & {\color{blue}list\_directory\_contents} & List files and directories with metadata        & \checkmark                       & \checkmark                        & \checkmark                         \\
		\midrule \multirow{3}{*}{\textbf{Code}}    & {\color{blue}extract\_source\_code}     & Extract source code by filename from APK        & \checkmark                       & \checkmark                        & \checkmark                         \\
		                                           & {\color{blue}search\_code\_patterns}    & Search keywords and patterns in source code     & \checkmark                       & \checkmark                        & \checkmark                         \\
		                                           & {\color{blue}enumerate\_source\_files}  & List important source files filtering libraries & \checkmark                       & \checkmark                        & \checkmark                         \\
		\midrule \multirow{6}{*}{\textbf{UI}}      & click\_ui\_element                      & Click UI elements (multi-strategy positioning)  &                                  & \checkmark                        &                                    \\
		                                           & input\_text\_field                      & Input text into UI text fields                  &                                  & \checkmark                        &                                    \\
		                                           & clear\_text\_field                      & Clear content from UI text fields               &                                  & \checkmark                        &                                    \\
		                                           & {\color{blue}verify\_element\_exists}   & Check UI element existence on current screen    &                                  & \checkmark                        & \checkmark                         \\
		                                           & {\color{blue}find\_ui\_elements}        & Find UI elements containing specific text       &                                  & \checkmark                        & \checkmark                         \\
		                                           & {\color{blue}capture\_ui\_layout}       & Capture XML hierarchy of current UI layout      &                                  & \checkmark                        & \checkmark                         \\
		\midrule \textbf{Log}                      & {\color{blue}search\_system\_logs}      & Search Android logcat using regex patterns      &                                  & \checkmark                        & \checkmark                         \\
		\midrule \multirow{3}{*}{\textbf{APK}}     & initialize\_gradle                      & Initialize project from benign template         &                                  & \checkmark                        &                                    \\
		                                           & build\_android\_package                 & Compile Android project to installable APK      &                                  & \checkmark                        &                                    \\
		                                           & install\_apk                            & Install APK file to connected Android device    &                                  & \checkmark                        &                                    \\
		\midrule \multirow{3}{*}{\textbf{Web}}     & initialize\_flask\_server               & Create Flask server directory structure         &                                  & \checkmark                        &                                    \\
		                                           & start\_web\_service                     & Start Flask development server                  &                                  & \checkmark                        &                                    \\
		                                           & stop\_web\_service                      & Terminate running Flask server process          &                                  & \checkmark                        &                                    \\
		\bottomrule
	\end{tabular}
	}
\end{table}

	\paragraph{\ding{197} PoC Planner}
	performs strategic planning. For each speculative vulnerability finding, it examines the flaw's
	characteristics, code context, and attack vectors to propose a step-by-step validation plan. As Table~\ref{tab:comprehensive_toolkit}
	shows, the \ding{197} \textit{Planner} only triggers ``read-only''\footnote{We use ``read-only''
	to refer to function calls that do not alter the runtime state of the target application inside
	the Android emulator. However, they may still manipulate the local analysis environment, such as
	creating a python script, executing code in it, and collecting results.} function calls, such as
	code navigation, file system analysis, and basic application \textit{control}. This ``read-only''
	design separates planning from execution and avoids unintended side effects. The \ding{197}
	\textit{Planner} may also attempt to further filter false positives after analyzing all supporting
	evidence. After a false positive is speculated, the \ding{197} \textit{Planner} assigns \textit{determine\_potential\_fp}
	tasks to the \ding{198} \textit{Task Executor} for concrete validation.

	\paragraph{\ding{198} Task Executor}
	implements the \ding{197} \textit{Planner}'s natural-language strategy as a concrete \ac{PoC}
	for speculative vulnerability findings. It sequentially executes the required steps using all eight
	toolkit categories ($29$ functions in total, cf. Table~\ref{tab:comprehensive_toolkit}),
	covering code execution, device control, file system, static analysis, UI interaction, log analysis,
	APK generation, and web server management. Acting as a multimodal agent, the \textit{Executor}
	processes both visual data (screenshots) and structural data (XML layouts) to navigate and
	interact with the victim application during PoC execution.

	\paragraph{\ding{199} Task Validator}
	provides independent verification of each \ac{PoC} outcome. It never accepts the \ding{198}
	\textit{Task Executor}'s self-reported success. Instead, after the \ding{198} \textit{Task
	Executor} runs a PoC, the \ding{199} \textit{Task Validator} uses its own observations and
	recomputation to judge whether the expected effect actually took place. \underline{\ding{199} Task Validator is the key to \tool{}'s success.}

	Similar to \ding{197}, the \ding{199} \textit{Task Validator} operates with ``read-only'' function
	calls. This gives \ding{199} wide visibility but prevents any modification of the application state
	(which may change the \ac{PoC} outcome, and we don't want that, cf. Table~\ref{tab:comprehensive_toolkit}).
	\begin{figure}[t]
		\centering
		\includegraphics[width=\linewidth]{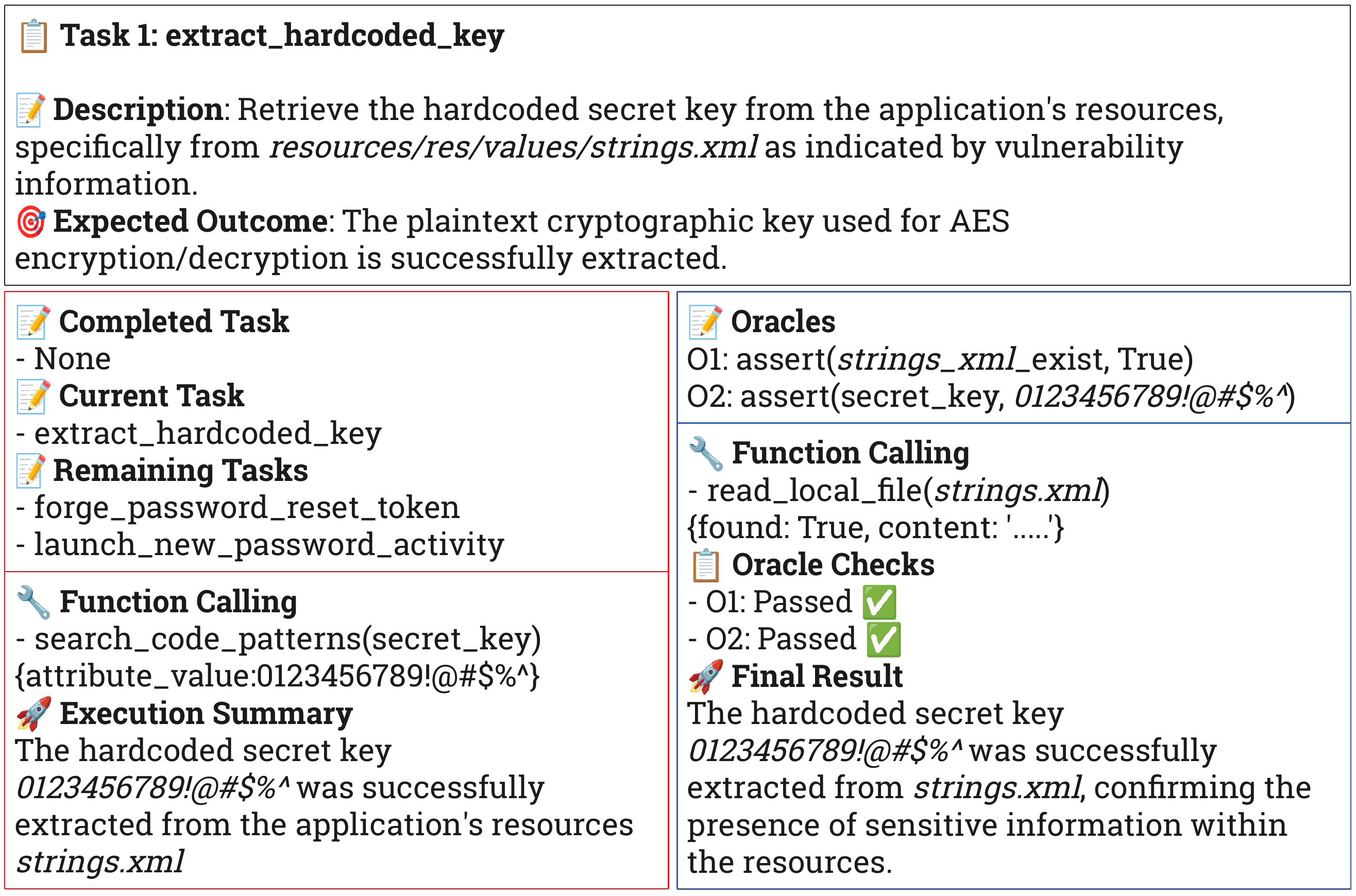}
		\caption{Hardcoded key extraction validation workflow. \tool{} searches for cryptographic
		key patterns within application resources and extracts a hardcoded AES key from the strings.xml
		file, with the \ding{199} \textit{Task Validator} verifying both file existence and key value
		through oracle-based validation.}
		\label{fig:example_task1}
	\end{figure}

	\begin{figure}[t]
		\centering
		\includegraphics[width=\linewidth]{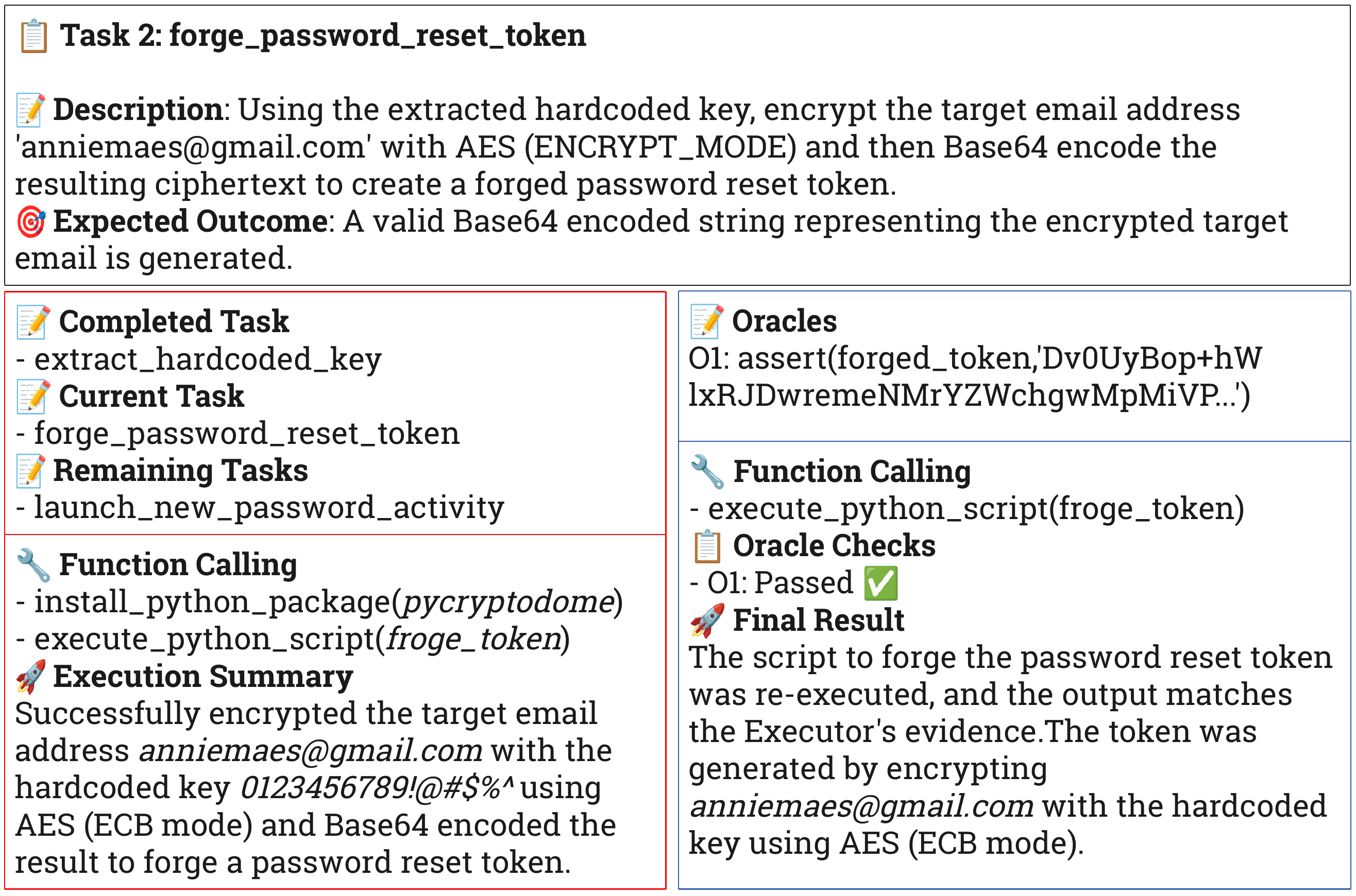}
		\caption{Cryptographic token forgery validation workflow. The \ding{198} \textit{Task
		Executor} uses the extracted AES key to encrypt a target email address and generate a Base64-encoded
		password reset token, while the \ding{199} \textit{Task Validator} independently verifies
		the cryptographic operations.}
		\label{fig:example_task2}
	\end{figure}

	\begin{figure}[t]
		\centering
		\includegraphics[width=\linewidth]{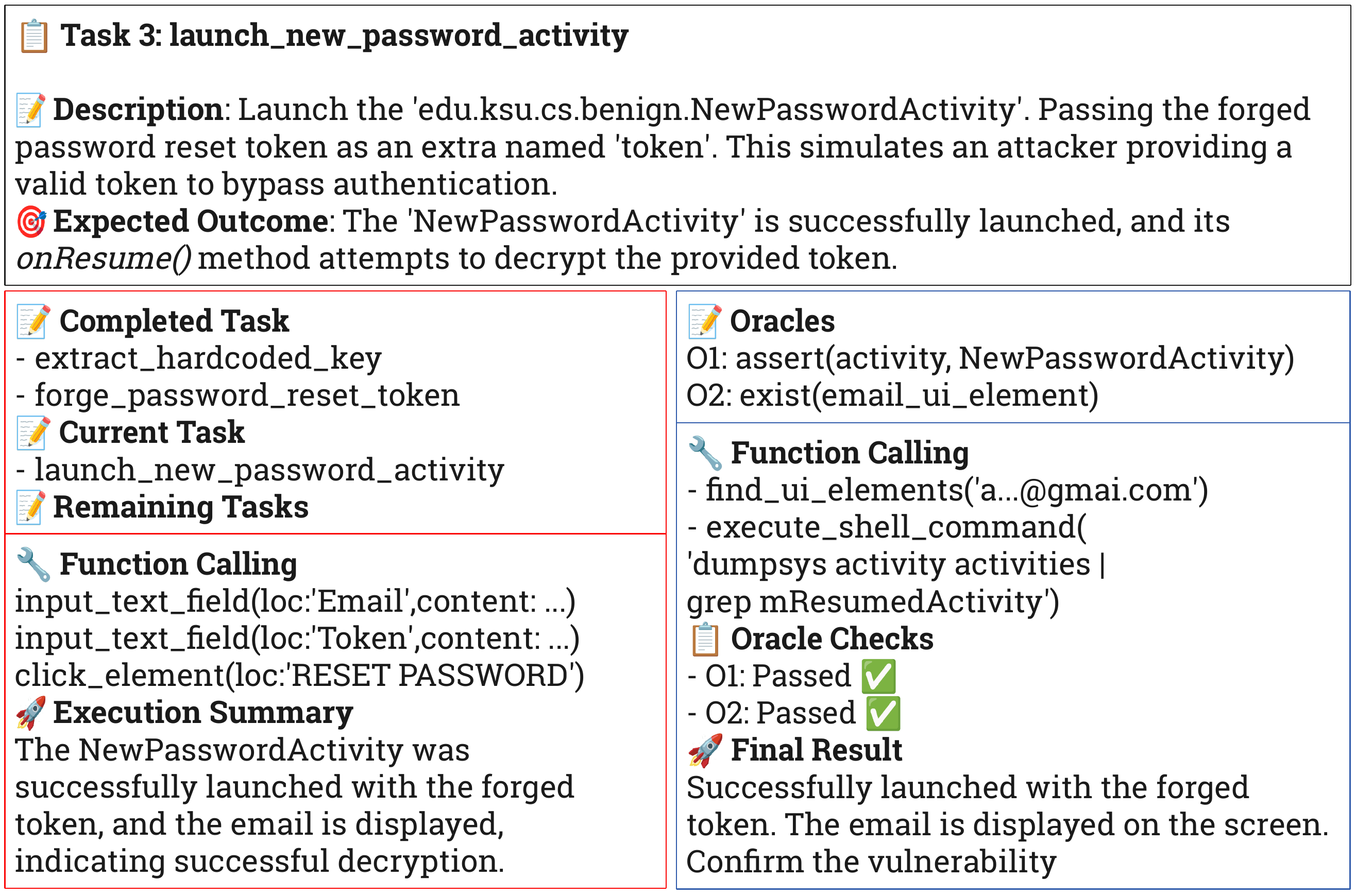}
		\caption{Authentication bypass validation through forged token injection. \tool{} launches
		the NewPasswordActivity with the crafted token, demonstrating successful authentication bypass
		as verified through UI element analysis and system activity monitoring.}
		\label{fig:example_task3}
	\end{figure}

	\begin{figure}[t]
		\centering
		\includegraphics[width=\linewidth]{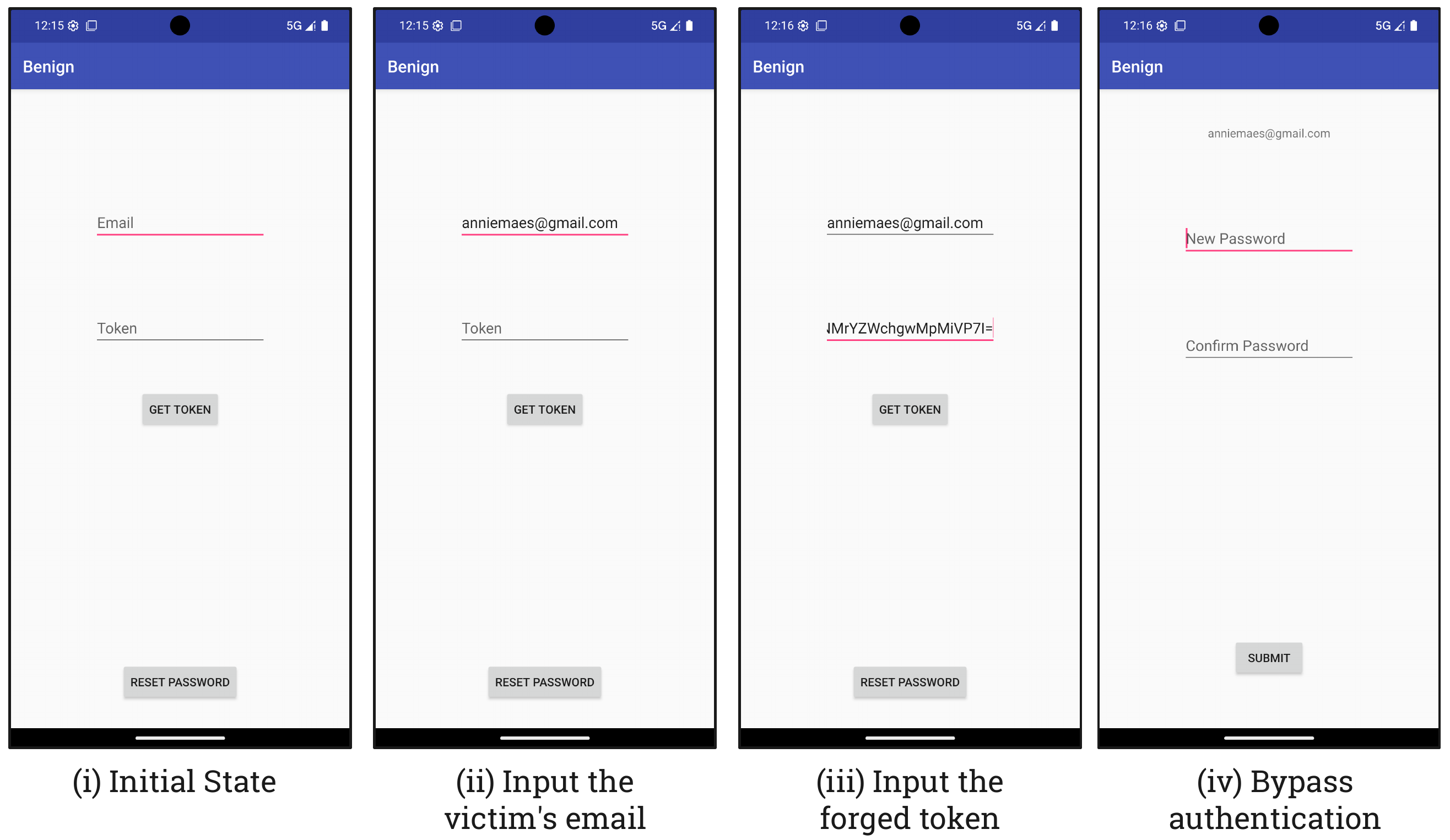}
		\caption{Progressive authentication bypass demonstration across four UI states: (i) initial password
		reset interface, (ii) target email input, (iii) forged token injection, and (iv) successful
		authentication bypass leading to password reset functionality access.}
		\label{fig:ui_change}
	\end{figure}

	\paragraph{Iterative Self-Correction}
	When execution fails at \ding{198} or validation rejects a success claim at \ding{199}, feedback
	is sent back to the \ding{197} \textit{PoC Planner}. The planner then inspects the failure,
	revises its strategy, and retries. This self-correction loop lets the system adapt and handle
	unexpected obstacles. The \ding{199} \textit{Task Validator} guides the process into one of three
	states: \textit{(i)} \textit{Continuation}, when a task is validated and the next task begins;
	\textit{(ii)} \textit{Re-planning}, when validation fails and the planner must revise its approach;
	or \textit{(iii)} \textit{Completion}, when all tasks pass validation and the workflow ends.

	\subsection{Vulnerability Oracle}
	\label{sec:verification_oracles}

	We highlight the oracle design as \underline{one of the core novelties} of \tool, since reliable
	validation is what ultimately separates speculative findings from actionable vulnerabilities.

	Unlike some security domains where vulnerability oracles can be defined with clear heuristics---for
	example, in blockchain security, monetary gain can serve as a definitive indicator~\cite{gervais2024ai}---Android
	vulnerability validation presents a much harder challenge. Applications expose diverse behaviors,
	and strict rules that guarantee zero false positives are difficult, if not impossible, to design.
	This limitation has long constrained fuzzers and static analysis tools in the Android ecosystem:
	they either over-approximate and flag benign behaviors or under-approximate and miss subtle
	flaws.

	\tool adopts a different strategy. Instead of relying on fixed heuristics, the \ding{199} \textit{Task
	Validator} uses \acp{LLM} to generate customized oracles for each vulnerability, case by case. Each
	claim from the \ding{198} \textit{Executor} is transformed into targeted verification questions.
	Guided by prior knowledge and the context of the current finding, the \ding{199} Validator
	designs oracles that independently test whether the claimed effect truly manifests. On a high level,
	an oracle involves the following steps:

	\begin{enumerate}
		\item \textbf{Parse the claim.} \ding{199} takes a single claim from the \ding{198} Executor,
			for example ``hardcoded AES key found,” or “admin screen reachable without login.''

		\item \textbf{State the expected effect.} \ding{199} writes the expected testable effect,
			for example ``the key decrypts file correctly,'' or ``AdminActivity opens while the session
			is unauthenticated.''

		\item \textbf{Design an oracle.} \ding{199} chooses one or more checks that can confirm or reject
			the effect. These checks use only “read only” tools from Table~\ref{tab:comprehensive_toolkit}.

		\item \textbf{Collect evidence.} \ding{199} run the selected tools to gather code, UI, logs,
			files, or network traces.

		\item \textbf{Decide.} \ding{199} compare the evidence with the expected effect. The oracle
			returns PASS or FAIL, plus a short rationale and artifacts for the report.
	\end{enumerate}

	We demonstrate \tool{}'s validation capabilities through a hardcoded cryptographic key
	vulnerability case from the Ghera benchmark~\cite{ghera}.

	\paragraph{Vulnerability Scenario}
	The target application implements a password reset verification interface where users provide their
	email address and click "GET TOKEN" to receive a verification token. After entering the correct
	token, users proceed to the password reset interface. The application uses the \textit{Cipher} API
	with AES encryption for token generation but stores the cryptographic key (\texttt{0123456789!@\#\$\%\^{}})
	directly in the \texttt{strings.xml} resource file. This enables attackers to extract the
	hardcoded key from the APK and forge authentication tokens for arbitrary users, bypassing security
	controls.

	\tool{} validates this vulnerability through three sequential tasks mirroring a real-world
	attack scenario. Figures~\ref{fig:example_task1}, \ref{fig:example_task2}, and \ref{fig:example_task3}
	demonstrate the validation workflow, while Figure~\ref{fig:ui_change} shows the UI changes
	during exploitation.

	\paragraph{Task 1: Hardcoded Key Extraction}
	In this initial validation step, \tool{} locates and extracts the hardcoded cryptographic key embedded
	within the application's resource files. As shown in Figure~\ref{fig:example_task1}, the system successfully
	identifies the AES key stored in the \texttt{strings.xml} configuration file. The \textit{Task
	Validator} independently verifies both the file's existence and the specific key value through
	oracle-based validation, confirming the presence of this critical security vulnerability.

	\paragraph{Task 2: Password Reset Token Forgery}
	Building on the extracted key, \tool{} demonstrates the exploitability of this vulnerability by
	forging authentication tokens for arbitrary users. Figure~\ref{fig:example_task2} illustrates how
	the system uses the hardcoded key to encrypt a target email address (\texttt{anniemaes@gmail.com})
	with AES encryption, generating a Base64-encoded password reset token. The validation process
	confirms that the generated token matches the expected cryptographic output and can successfully
	impersonate legitimate users.

	\paragraph{Task 3: Authentication Bypass Validation}
	The final validation step proves that the forged tokens can bypass the application's authentication
	mechanisms in practice. As demonstrated in Figure~\ref{fig:example_task3}, \tool{} launches the
	password reset functionality using the crafted token, successfully deceiving the application into
	accepting it as legitimate. Figure~\ref{fig:ui_change} shows the complete UI progression across
	four states, documenting how the attack grants unauthorized access to password reset functionality,
	thereby confirming the end-to-end exploitability of the vulnerability.

	\section{Evaluation}
	\label{sec:evaluation}

	In this section, we conduct experiments to evaluate \tool's performance, cost effectiveness, and
	capability to discover zero-day vulnerabilities in the wild. We compare \tool against \ac{SOTA}
	static analysis tools and discss the quality of \tool's generated \acp{PoC} across multiple dimensions.

	\subsection{Dataset Selection}
	\label{subsec:dataset}

	Our experiments are conducted on two datasets.

	\begin{itemize}
		\item The \textit{Ghera} benchmark dataset contains 60 vulnerable Android APKs spanning
			eight categories—CRYPTO, ICC, NETWORKING, NON-API, PERMISSION, STORAGE, SYSTEM, and WEB—and
			serves as our primary testbed for both detection and exploit generation evaluation.

		\item For real-world vulnerability discovery, we use APKs from AndroZoo~\cite{allix2016androzoo}
			released between January 2024 and July 2025. To address model context limits during
			detection, we restrict to APKs under~$5$MB, yielding~$169$ distinct samples: $126$ from
			play.google.com, $38$ from VirusShare, and $5$ from other platforms.
	\end{itemize}

	To reduce potential data leakage and model memorization, we remove all textual descriptions or
	semantic information about the vulnerabilities from our dataset. In addition, all application-specific
	identifiers, such as filenames and APK names, are replaced with generic placeholders.

	\subsection{Experimental Setup}
	\label{subsec:model}

	We describe below the setup used for all experiments.

	\paragraph{Computational Environment.}
	All experiments run on an Apple M4 system with 24GB RAM and 1TB SSD storage (macOS 26.0). The Android
	testing setup uses emulator version 35.6.11.0, adb 1.0.41 (build 36.0.0-13206524), and Android API
	level 36. Experiments are conducted with internet connectivity for \ac{LLM} API access, with a 300-second
	timeout per call to ensure reliable execution.

	\paragraph{Comparison Tool Selection.}
	We select tools based on three criteria: \textit{(i)} \emph{APK compatibility}, meaning the tool
	natively supports compiled Android Package files without requiring source code; \textit{(ii)}
	\emph{broad vulnerability coverage}, spanning multiple categories rather than focusing on a
	narrow class of flaws; and \textit{(iii)} \emph{active maintenance and adoption}, ensuring the
	tool is available and representative of current practice. These three selection criteria favor static
	tools. Dynamic analysis is particularly challenging for Android because of its multimodal input
	space -- UI events, Intents, filesystem access, and database queries -- which makes general-purpose
	DAST difficult to apply~\cite{10504267}. Existing Android DAST tools are usually specialized (e.g.,
	fuzzers for media frameworks or native libraries~\cite{peixoto2024fuzzing,blanda2015fuzzing}) or
	UI fuzzers aimed at functional testing~\cite{demissie2025vlmfuzz,liu2024visiondroid,xiong2023empirical},
	rather than security vulnerabilities. In this paper, we evaluate three widely used \ac{SAST}
	tools that cover a broad range of Android vulnerabilities: MobSF~\cite{mobsf}, APKHunt~\cite{apkhunt},
	and Trueseeing~\cite{trueseeing}. As of our paper submission, their features are summarized in Table~\ref{tab:tool_comparison}.
	\begin{table}[t]
	\centering
	\scriptsize
	\caption{Comparison of static tools evaluated in this study}
	\label{tab:tool_comparison} \resizebox{0.9\columnwidth}{!}{%
	\begin{tabular}{ccccc}
		\toprule Tool                & Stars & Commits & Version & Language \\
		\midrule MobSF~\cite{mobsf}  & 19k+  & 2,000+  & v4.3.2  & Python   \\
		APKHunt~\cite{apkhunt}       & 800+  & 100+    & v1.0.0  & Go       \\
		Trueseeing~\cite{trueseeing} & 60+   & 850+    & v2.2.7  & Python   \\
		\bottomrule
	\end{tabular}%
	}
\end{table}

	\paragraph{LLM Selection.}
	\label{para:llm_selection} To cover a representative range of performance and cost, we evaluate four
	commercial and open-source multimodal \acp{LLM}: OpenAI o3 (o3 2025-04-16), Gemini 2.5 Pro (gemini-2.5-pro),
	Gemini 2.5 Flash (gemini-2.5-flash), and GPT oss (gpt-oss-120b). At evaluation time, the
	advertised prices per million input and output tokens are 2 and 8 USD, 1.25 and 10 USD, 0.30 and
	2.50 USD, and 0.10 and 0.50 USD, respectively. To keep reasoning capacity consistent across models,
	we set \textit{thinkingBudget} to 24576 for gemini-2.5-flash and gemini-2.5-pro, and use the
	\textit{Medium} thinking intensity for o3 and gpt-oss-120b.

	\subsection{Detection Performance}
	\begin{table}[t]
	\centering
	\caption{Summary of \tool{} detection results on the Ghera benchmark and real-world APKs.
	\textbf{APK} is the number of applications analyzed. \textbf{LoC} is the average lines of code per
	application. \textbf{Alert} is the number of vulnerabilities initially reported by \tool{}. \textbf{Stage}
	distinguishes Agentic Vulnerability Discovery and Agentic Vulnerability Validation. \textbf{TP}
	and \textbf{FP} are true and false positives. \textbf{B\_TP} is the number of Ghera ground-truth
	vulnerabilities detected (out of 60). Validation consistently reduces false positives (e.g., Ghera:
	8→1, real-world APKs: 32→3) while preserving most true positives.}
	\label{tab:detection_results_summary} \resizebox{\columnwidth}{!}{
	\begin{tabular}{@{}cccccccc@{}}
		\toprule \centering Dataset                     & \centering APK                  & \centering LoC                      & \centering Alert                & \centering Stage                 & \centering TP & \centering FP & \centering B\_TP \tabularnewline \midrule \multirow{2}{*}{\centering Ghera} & \multirow{2}{*}{\centering 60} & \multirow{2}{*}{\centering 46,100} & \multirow{2}{*}{\centering 82} & Agentic Vulnerability Discovery & 75 & 7 & 47 \\
		                                                &                                 &                                     &                                 & Agentic Vulnerability Validation & 51            & 1             & 34                                                                           \\
		\midrule \multirow{2}{*}{\centering Production} & \multirow{2}{*}{\centering 169} & \multirow{2}{*}{\centering 115,696} & \multirow{2}{*}{\centering 136} & Agentic Vulnerability Discovery  & 104           & 32            & --                                                                           \\
		                                                &                                 &                                     &                                 & Agentic Vulnerability Validation & 57            & 3             & --                                                                           \\
		\bottomrule
	\end{tabular}
	}
\end{table}

	To ensure fairness, each LLM model is executed exactly once per APK. We collect all outputs
	directly, without cherry-picking or repeated attempts. For static tools, we use their default analysis
	pipelines. Ground-truth validation is performed manually by both authors, with cross-checking to
	confirm whether each tool correctly identifies the vulnerabilities documented in Ghera. Table~\ref{tab:vuln_detect_summary}
	in the Appendix presents the full detection results.

	Note that Ghera labels APKs with specific vulnerabilities but does not guarantee that other
	flaws are absent. For example, one APK labeled with \textit{BlockCipher-ECB-InformationExposure}
	also contains a \textit{Hard-coded Cryptographic Key}. Exhaustively identifying all vulnerabilities
	is non-trivial. To reduce subjective bias, we benchmark strictly against the vulnerabilities explicitly
	documented in Ghera.

	\paragraph{Reported Findings Volume.}
	We observe a clear gap in the number of findings between \ac{SAST} tools and our LLM-based analyzers.
	MobSF reports 5,654 findings, while o3 produces only 116. The large output of \ac{SAST} tools
	comes from broad rules that flag application issues, third-party libraries, SDK usage, and even coding
	style. This adds noise and makes audits harder for analysts.

	\begin{figure}[tb]
		\centering
		\includegraphics[width=\linewidth]{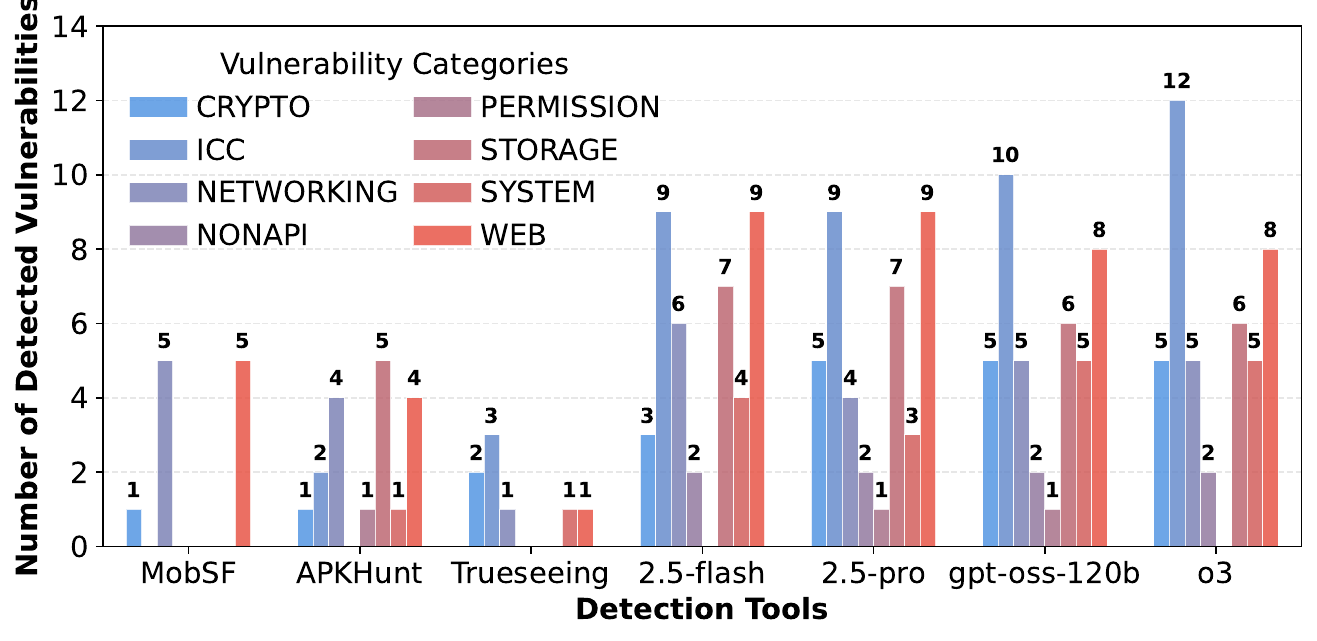}
		\caption{Detection performance across vulnerability categories. Each bar shows the number of
		detected vulnerabilities by category for each tool, with different colors representing
		different vulnerability categories.}
		\label{fig:stacked_detection}
	\end{figure}

	\paragraph{Detection Capability.}
	Figure~\ref{fig:stacked_detection} shows the number of vulnerabilities detected per tool, broken
	down by category. LLM-based analyzers achieve substantially higher recall than \ac{SAST} tools.
	o3 detects 43 vulnerabilities (71.7\%), while APKHunt detects 18 (30.0\%) and MobSF only 11 (18.3\%),
	despite MobSF’s much larger output volume. This contrast underscores the gap between concise LLM
	results and the noisy but shallow coverage of traditional static analysis.

	LLM analyzers also achieve broader category coverage. They consistently detect more vulnerabilities
	in CRYPTO, ICC, and STORAGE, categories where \ac{SAST} tools often fail. In comparison, \ac{SAST}
	tools display narrow specialization: APKHunt and Trueseeing each identify unique cases, but
	their coverage is narrow. APKHunt detects four CRYPTO vulnerabilities that no other \ac{SAST}
	tool reports, while Trueseeing is the only tool to flag three NONAPI issues and a single WEB case.
	Outside these isolated hits, their contribution is minimal, highlighting a lack of
	generalization beyond their targeted rules. Variance in performance further highlights this divide.
	\ac{SAST} tools diverge widely because their rules are tied to specific detection oracles, while
	LLM analyzers maintain a more balanced profile across categories. Analysis shows that LLMs subsume
	and extend traditional detection. The three LLMs—gemini 2.5 flash, gemini 2.5 pro, and
	o3—collectively capture all 27 vulnerabilities found by the three \ac{SAST} tools, while also
	uncovering 20 additional vulnerabilities that no \ac{SAST} tool detects (marked with $^{*}$ in
	Table~\ref{tab:vuln_detect_summary}). This demonstrates that LLM-based analysis not only
	preserves static-analysis strengths but also extends detection into previously unreachable
	ground. Despite these gains, 13 vulnerabilities remain undetected by any tool, a limitation discussed
	in Section~\ref{sec:limitations}.

	\subsection{Aggregation Effectiveness}

	The aggregation process collects and normalizes the heterogeneous outputs from all seven
	detection tools (three \ac{SAST} tools and four LLM-based analyzers) into a unified \textit{StandardizedVulnerability}
	format. For each of the 60 vulnerabilities in the Ghera benchmark, every aggregation model
	receives the complete set of outputs from all tools as input, regardless of whether individual tools
	flagged the vulnerability. We evaluate four LLM models as aggregation engines: gemini-2.5-flash (G2.5F),
	gemini-2.5-pro (G2.5P), o3, and gpt-oss-120b (oss). We record the outputs from each model and perform
	manual verification to check whether the aggregated results match the ground-truth vulnerabilities.

	\paragraph{Coverage Analysis.}
	Table~\ref{tab:vuln_detect_summary} in the Appendix shows that aggregation improves detection
	coverage. By combining the outputs from all seven tools, each of the four aggregation models
	identifies 47 of the 60 ground-truth vulnerabilities, achieving 78.3\% recall. This is a clear gain
	over the best individual tool, o3, at 71.7\%. The results highlight the complementary strengths of
	different detection approaches.

	\begin{figure}[t]
		\centering
		\includegraphics[width=0.85\linewidth]{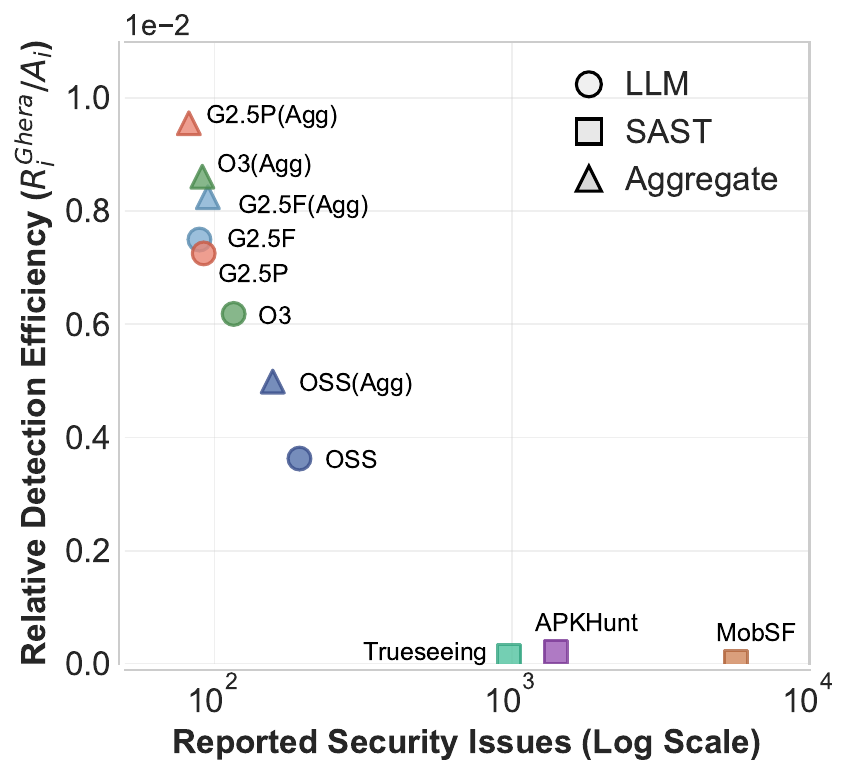}
		\caption{Detection efficiency analysis across tool categories. The scatter plot shows reported
		security issues (x-axis, log scale) versus relative detection efficiency (y-axis,
		$R_{i}^{Ghera}/A_{i}$ in scientific notation). Circles represent LLM-based analyzers, squares
		represent SAST tools, and triangles represent aggregated results. The ``(Agg)'' suffix indicates
		aggregated configurations merging multiple tool outputs. Each tool uses consistent color
		coding for individual and aggregated versions. Higher efficiency values indicate better balance
		between detection coverage and warning / speculative vulnerability finding volume, with LLM-based
		tools substantially outperforming SAST approaches.}
		\label{fig:detection_efficiency}
	\end{figure}

	\paragraph{Consolidation Effectiveness.}
	While all four aggregation models achieve identical top-line recall performance, they exhibit
	notable differences in their consolidation strategies. For example, in analyzing \textit{SQLite-RawQuery-SQLInjection}
	vulnerabilities, gemini-2.5-flash and gemini-2.5-pro consolidate findings into 4 distinct issues,
	whereas o3 reports only 2. Conversely, for \textit{ConstantKey-ForgeryAttack}, o3 provides a
	more granular report with 2 issues compared to 1 from the other models. Notably, gpt-oss-120b
	demonstrates suboptimal consolidation performance, suggesting a higher tendency toward hallucination
	and less effective noise filtering compared to other models. Overall, gemini-2.5-flash, gemini-2.5-pro,
	and o3 demonstrate strong performance in producing concise, actionable final reports.

	\paragraph{Detection Efficiency Analysis.}
	Security practitioners must balance detection coverage against analysis overhead. Traditional
	tools often produce thousands of noisy findings, complicating assessment and inflating audit costs.
	To capture this trade-off, we define a detection efficiency metric that relates recall to warning
	volume. Formally, the efficiency of tool $i$ is:
	\begin{equation}
		E_{i}= \frac{R_{i}^{Ghera}\times V_{total}}{A_{i}}
	\end{equation}
	where $R_{i}^{Ghera}$ is the recall of tool $i$ on the labeled vulnerabilities in the Ghera
	benchmark, $A_{i}$ is the number of warnings / speculative vulnerability findings it reports,
	and $V_{total}$ represents the true (but unknown) total number of vulnerabilities across the
	dataset. We only know $V_{total}\geq 60$, since 60 vulnerabilities are labeled.

	Because $V_{total}$ is constant across all tools, relative comparisons depend only on the ratio $R
	_{i}^{Ghera}/A_{i}$. Intuitively, this ratio reflects the expected number of true findings per warning,
	scaled by recall. We use this proxy because it emphasizes the trade-off practitioners face:
	higher recall increases the chance of covering real vulnerabilities, while excessive warnings increase
	audit cost. Although other formulations (e.g., precision or F1-score) are possible, this measure
	directly captures the efficiency of converting warnings into actionable detections.

	Figure~\ref{fig:detection_efficiency} compares efficiency scores across tools. Three trends
	stand out. \textbf{LLM advantage:} LLM analyzers deliver higher recall (66.7–71.7\%) while producing
	far fewer warnings (89–193) than \ac{SAST} tools (978–5,654). \textbf{Aggregation effectiveness:}
	Combining outputs achieves the highest efficiency, reaching 78.3\% recall with only 82–157
	speculative vulnerability findings, highlighting the complementary strengths of different methods.
	\textbf{SAST limitations:} Static tools remain inefficient. MobSF, for example, generates over 5,000
	warnings yet achieves only 18.3\% recall.

	\subsection{Validation Effectiveness}
	\label{subsec:validation_effectiveness}

	We validate the 82 speculative vulnerability findings produced by gemini-2.5-pro in the
	aggregation stage. For clarity, we exclude two groups of findings from evaluation: \textit{(i)}
	two findings from APKs that could not be installed on the emulator (\textit{WebViewAllowContentAccess-UnauthorizedFileAccess}
	and \textit{WebViewLoadDataWithBaseUrl-UnauthorizedFileAccess}); and \textit{(ii)} seventeen
	findings corresponding to vulnerabilities outside \tool{}’s current validation scope (see Section~\ref{sec:limitations}).

	The remaining 63 (82-2-17) speculative vulnerability findings, listed in Table~\ref{tab:poc_summary}
	in the Appendix, undergo manual review and classification. This process identifies 56 true positives
	and 7 false positives (marked in yellow in the table). To ensure accuracy, verification is cross-checked
	by multiple security researchers. We then evaluate these 63 findings under two model
	configurations for the three-agent system. The first configuration uses (gemini-2.5-pro, gemini-2.5-flash,
	gemini-2.5-flash) for the (planner, executor, validator) roles, representing a cost-optimized
	setup. The second configuration uses (gemini-2.5-pro, gemini-2.5-pro, gemini-2.5-pro) across all
	roles, representing a performance-optimized setup with consistent model strength throughout the
	pipeline. For simplicity, we use \textbf{Mixed Configuration} and \textbf{Unified Configuration}
	to refer to these two setups respectively in subsequent sections. Validation outcomes are categorized
	with standardized symbols: \ding{72} = true positive validated with a complete proof-of-concept;
	\ding{75} = false positive correctly identified and filtered; $\otimes$ = true positive
	misclassified as false positive; $\odot$ = false positive misclassified as true positive;
	$\times$ = execution terminated due to technical errors; $\bullet$ = validation stopped after reaching
	the maximum iteration limit.

	\begin{figure}[t]
		\centering
		\includegraphics[width=0.9\linewidth]{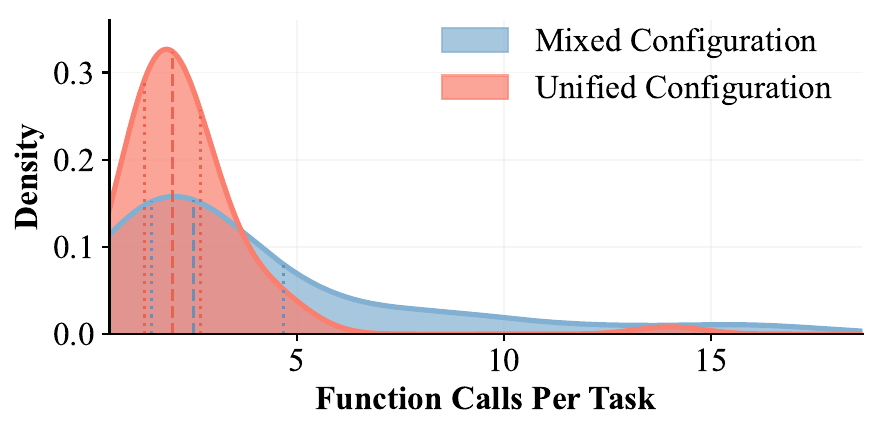}
		\caption{Kernel density estimation of function calls per task distribution across both model
		configurations, showing density peaks around 2--4 function calls per task.}
		\label{fig:function_calls_per_task_kde}
	\end{figure}

	\paragraph{Task Executor Performance.}
	Function calling success rates reach 92.5\% (927/1002) for mixed configuration and 95.4\% (649/680)
	for unified configuration, with the majority of failures concentrated in APK generation and web
	server management functions. These failures primarily result from environment-specific challenges
	including Gradle build system configuration conflicts, network port allocation issues, and Android
	emulator state management complexities. Figure~\ref{fig:function_calls_per_task_kde} shows the
	function calls per task distribution.
	\begin{figure*}[t]
		\centering
		\includegraphics[width=0.24\textwidth]{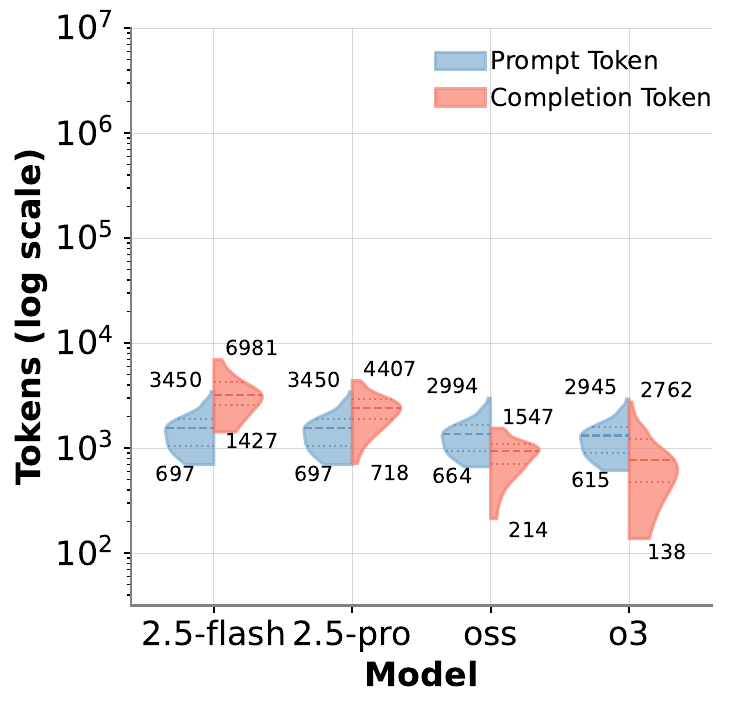}
		\hfill
		\includegraphics[width=0.24\textwidth]{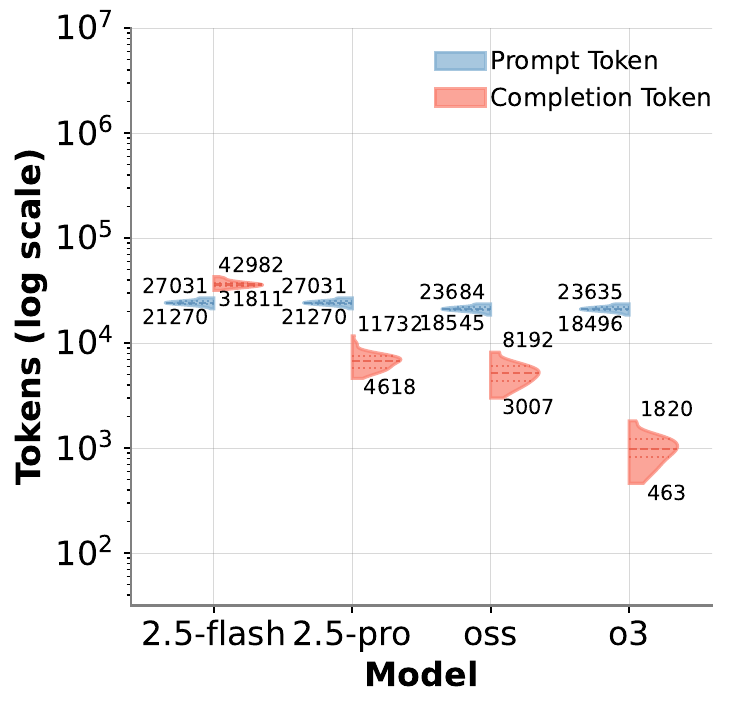}
		\hfill
		\includegraphics[width=0.24\textwidth]{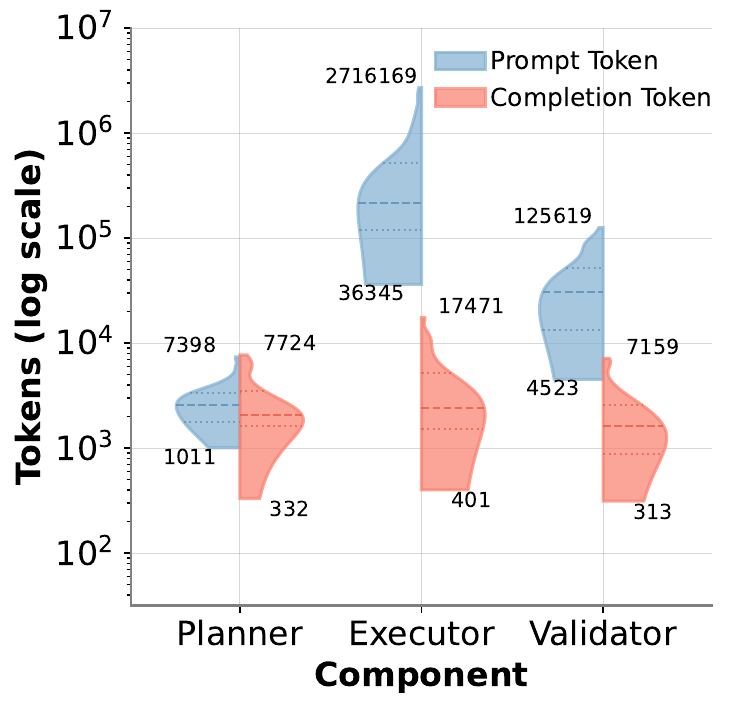}
		\hfill
		\includegraphics[width=0.24\textwidth]{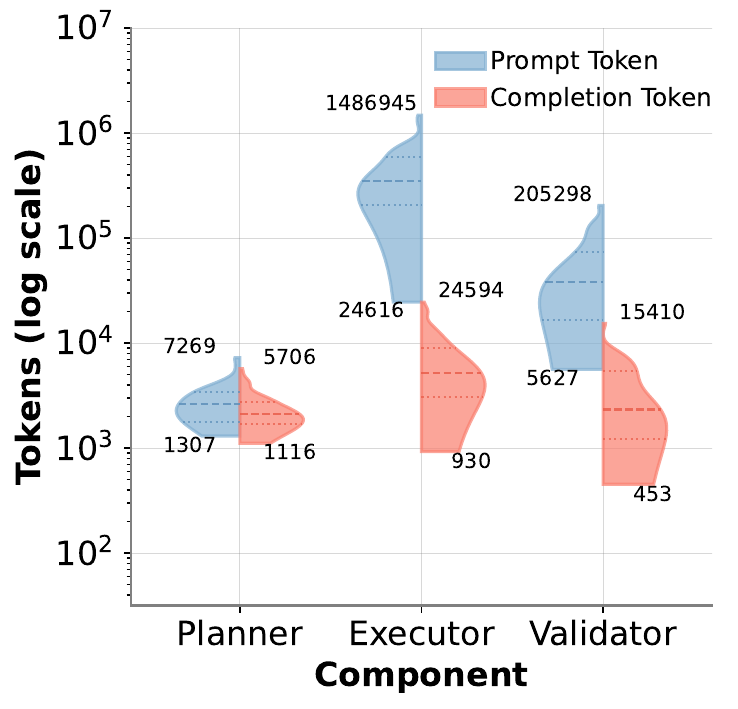}
		\\[0.2em]
		\makebox[0.24\textwidth]{(a) Detection Phase}
		\hfill
		\makebox[0.24\textwidth]{(b) Aggregation Phase}
		\hfill
		\makebox[0.24\textwidth]{(c) Mixed Configuration}
		\hfill
		\makebox[0.24\textwidth]{(d) Unified Configuration}
		\caption{Token consumption analysis across different phases and configurations. Green areas represent
		prompt tokens, pink areas represent completion tokens. (a) Individual model performance during
		vulnerability detection, (b) Token usage during multi-tool result aggregation, (c) Component-wise
		analysis with Planner using gemini-2.5-pro and Executor/Validator using gemini-2.5-flash, (d)
		All components using gemini-2.5-pro.}
		\label{fig:token_analysis}
	\end{figure*}

	The function calls per task distribution indicates that 80\% of tasks complete within 2-4 function
	calls, with unified configuration demonstrating reduced function call requirements (average 2.92
	vs 4.49 calls per task). This represents a +35\% reduction in API overhead.

	\paragraph{Validator Performance.}
	The \textit{Task Validator} processes 223 task execution claims from the \textit{Task Executor}
	in mixed configuration and 233 claims in unified configuration, achieving validation pass rates of
	87.4\% (195/223) and 95.3\% (222/233) respectively. The validator identifies and filters 28
	false positive execution claims in mixed configuration and 11 false positive claims in unified
	configuration, corresponding to hallucination rates of 12.6\% and 4.7\%. These results quantify
	the validator's role in maintaining execution reliability and preventing the propagation of
	erroneous claims through the validation pipeline.

	The validator's hint generation mechanism operates through dynamic oracle construction, providing
	specific failure context when validation fails. Analysis of retry sequences shows that 78\% of initially
	failed tasks achieve successful validation after incorporating validator feedback, demonstrating
	the iterative refinement capability of the multi-agent system.

	\paragraph{Overall Validation Results.}
	Among the 82 speculative vulnerability findings from aggregation, our analysis identifies 56
	true positives, 7 false positives, 17 out-of-scope cases, and 2 unable-to-install instances.
	Mixed configuration achieves validation of 46 out of 75 actionable cases (61.3\%), while unified
	configuration reaches 51 out of 75 cases (68.0\%), representing a +6.7 percentage point improvement.
	For end-to-end effectiveness based on the 60 labeled vulnerabilities in Ghera, those marked with
	asterisks (*) in Table~\ref{tab:poc_summary} in the Appendix correspond to ground truth labels.
	Mixed configuration validates 31 labeled vulnerabilities (51.7\%), while unified configuration
	achieves 34 (56.7\%). This +5.0 percentage point improvement demonstrates the impact of consistent
	model capability across validation agents, confirming the importance of model selection in
	autonomous security analysis.

	\subsection{Cost and Efficiency Analysis}

	To evaluate the practical deployment feasibility of \tool{}, we analyze computational costs and efficiency
	across different system configurations and operational phases.

	\paragraph{Token Consumption Analysis.}
	Token consumption exhibits distinct scaling patterns across operational phases. During individual
	vulnerability detection (Figure~\ref{fig:token_analysis}a), models consume 1,322-1,552 prompt tokens
	and 778-3,523 completion tokens per APK. The aggregation phase (Figure~\ref{fig:token_analysis}b)
	requires 19,074-24,477 prompt tokens and 986-36,624 completion tokens, representing a 15-20x
	increase due to processing outputs from seven detection tools. In the multi-agent validation pipeline,
	the Executor component dominates consumption with median prompt tokens of 222,733 (mixed configuration,
	Figure~\ref{fig:token_analysis}c) and 350,667 (unified configuration, Figure~\ref{fig:token_analysis}d),
	while \textit{Planner} and \textit{Validator} components require substantially fewer resources.
	This distribution indicates that 70-80\% of computational costs concentrate in the exploitation generation
	phase, suggesting optimization opportunities.

	\paragraph{Economic Cost Assessment.}
	Based on model pricing in our LLM selection criteria~\ref{para:llm_selection}, we estimate per-vulnerability
	validation costs across configurations. Detection-only costs range from \$0.003-0.029 per APK (o3),
	\$0.0004-0.001 per APK (gpt-oss-120b), to \$0.002-0.014 per APK (Gemini variants). Aggregation increases
	costs to \$0.04-0.33 per APK for gpt-oss-120b, \$0.06-0.66 per APK for gemini-2.5-flash, \$0.26-0.61
	per APK for gemini-2.5-pro, and \$0.84-3.35 per APK for o3. Full validation pipeline costs vary significantly:
	mixed configuration averages \$0.59-4.23 per vulnerability (median \$1.77), while unified gemini-2.5-pro
	configuration ranges \$4.81-26.85 per vulnerability (median \$8.94). Despite higher per-token
	costs, unified configuration achieves 82\% reduction in \textit{Executor} token usage, though
	this translates to only marginal cost savings due to gemini-2.5-pro's higher pricing structure.

\begin{figure}[t]
		\centering
		\includegraphics[width=0.9\linewidth]{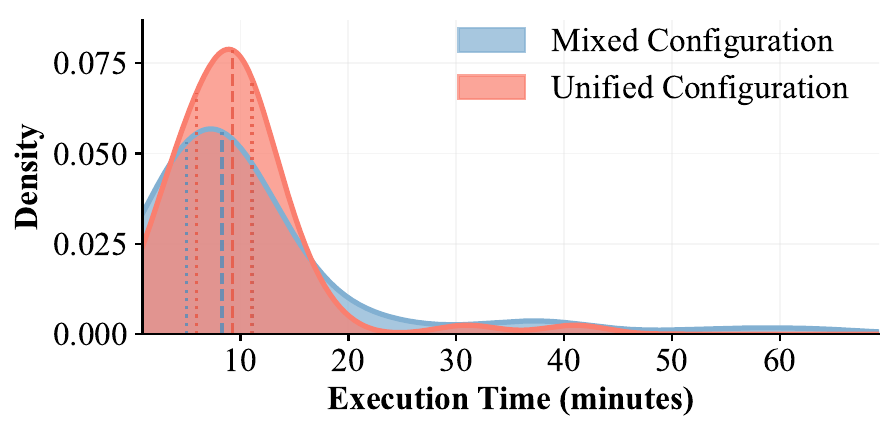}
		\caption{Kernel density of execution time distribution during vulnerability validation
		across both model configurations.}
		\label{fig:execution_time_kde}
	\end{figure}

	\paragraph{Efficiency Analysis.}
	We analyze \tool{}'s time overhead. During Agentic Vulnerability Discovery, efficiency
	differences primarily stem from varying model request times since this phase involves code-level
	analysis without dynamic APK interaction. We focus on Agentic Vulnerability Validation time overhead.
	Figure~\ref{fig:execution_time_kde} presents execution time distributions across both
	configurations. Mixed configuration achieves mean execution time of 11.14 minutes (median 8.29
	minutes, $\sigma$ = 11.70), while unified configuration demonstrates 9.62 minutes mean (median
	9.24 minutes, $\sigma$ = 6.44). Both configurations complete 75\% of vulnerabilities within 11
	minutes, indicating comparable baseline throughput. However, temporal outliers extend to 63 minutes
	(mixed) and 41 minutes (unified), representing $<$5\% of validation cases requiring extended
	analysis. The unified configuration exhibits 45\% reduced variance and 35\% lower maximum execution
	time due to reduced function call requirements, enabling validation completion with fewer
	execution steps.

	\section{Real-World Vulnerability Detection}
	\label{sec:real_world_vuln_detection}

	To evaluate \tool{}'s effectiveness on real-world production applications, we conduct
	experiments on a real-world dataset comprising 169 APKs with an average of 115,696 lines of code
	per application. The experimental configuration employs (i) two gemini-2.5-pro models as detectors
	with gemini-2.5-flash as aggregator, and (ii) gemini-2.5-flash as planner, executor, and
	validator for the validation pipeline. After excluding 9 APKs that cannot be installed on the Android
	emulator, we analyze the remaining 160 APKs.

	\paragraph{Detection Results.}
	Table~\ref{tab:detection_results_summary} presents the detection results. We apply \tool{} to scan
	the 160 APKs. During the Agentic Vulnerability Discovery phase, \tool{} reports 136 speculative
	vulnerabilities. In the Agentic Vulnerability Validation phase, \tool{} successfully reports 60 validated
	vulnerability findings and filters out 29 false positive findings. Through manual review, we further
	identify that only 3 out of the 60 validated vulnerability findings are false positives,
	resulting in a false positive rate of 5.0\% (3/60).

	\paragraph{Vulnerability Classification.}
	Table~\ref{tab:vuln_classification} analyzes the distribution of true positive vulnerability
	types across discovery and validation phases. Agentic Vulnerability Validation demonstrates high
	effectiveness for Data Exposure vulnerabilities, validating 77.3\% (17/22) of speculative findings.
	Code Injection vulnerabilities also show strong validation rates at 80.0\% (12/15). In contrast,
	Network Communication vulnerabilities exhibit low validation rates at 4.3\% (1/23). Similar to the
	Ghera dataset findings, network-related vulnerabilities involve man-in-the-middle attacks, certificate
	manipulation, and network traffic monitoring that require infrastructure beyond the Android
	emulator capabilities that \tool{} currently lacks.

	\paragraph{Responsible Disclosure.}
	Following security research practices, we responsibly disclose identified vulnerabilities through
	bug bounty programs, official guidelines, or direct contact with developers. Confirmed cases are
	documented and reported via proper channels, with sufficient time for remediation before public release.
	The findings span cryptographic, input validation, and access control flaws, demonstrating \tool{}’s
	broad detection capabilities in real-world Android apps. Full details will be shared after
	remediation.

	\begin{table}[t]
	\caption{Classification of True Positive (TP) Test Cases by Vulnerability Type. Speculative refers
	to speculative vulnerability findings, and Validated refers to validated vulnerability findings.}
	\scriptsize \label{tab:vuln_classification}
	\centering
	\begin{tabular*}{0.9\linewidth}{@{\extracolsep{\fill}}lcc}
		\toprule Type                 & Speculative & Validated \\
		\midrule Authorization        & 19          & 9         \\
		Network Communication         & 23          & 1         \\
		Code Injection                & 15          & 12        \\
		WebView                       & 13          & 9         \\
		Data Exposure                 & 22          & 17        \\
		Inter-Component Communication & 12          & 9         \\
		\midrule Total                & 104         & 57        \\
		\bottomrule
	\end{tabular*}
\end{table}

	\section{Discussion}

	\paragraph{Do We Still Need SAST Tools?}
	Despite \ac{LLM}-based analyzers achieving superior vulnerability coverage compared to
	individual \ac{SAST} tools, traditional static analysis retains critical value within \tool{}'s architecture.
	\textit{(i) Exploit Path Guidance}: Static tools excel at providing code-line granular
	vulnerability localization and data flow analysis that enables precise source-to-sink path
	identification. This information, when aggregated into standardized reports, assists the \ac{PoC}
	Planner in synthesizing targeted attack vectors through precise entry point identification. \textit{(ii)
	Hallucination Filtering}: Call graph analysis mitigates \ac{LLM} false positives when context
	windows approach model capacity limits, preventing phantom function invocation claims that lead to
	validation overhead. Consider the following anonymized code fragment from our real-world
	evaluation dataset (Figure~\ref{lst:path_traversal}):

	\begin{figure}[t]
		\adjustbox{scale=0.95}{%
		\begin{minipage}{\linewidth}\begin{lstlisting}[language=Java,
		moredelim={[is][\color{red!60!black}]{@}{@}}]
public static boolean isClassPreloadingAppEnabled(Context context) {
    File fileA01 = @AnonymousClass000@.A01(context);
    boolean zExists = @fileA01@.exists();
    @fileA01@.delete();
    return zExists;
}
public static boolean isClassPreloadingEnabled(Context context, String str) {
    File fileA01 = @AnonymousClass000@.A01(context, str);
    boolean zExists = @fileA01@.exists();
    @fileA01@.delete();
    return zExists;
}
		\end{lstlisting}\end{minipage} }
		\caption{A path traversal vulnerability reported in discovery stage but eliminated in
		validation stage.}
		\label{lst:path_traversal}
	\end{figure}

	Our \ac{LLM} analyzer identified this as a \textit{Path Traversal} vulnerability due to unsanitized
	string parameters directly constructing file paths, potentially leading to deleting critical
	databases or shared preferences files. However, call graph analysis revealed these methods lack
	both internal invocation and external exposure, rendering the speculative vulnerability finding a
	false positive despite structural vulnerability presence. While our validation framework
	correctly classified this as \textit{FP}, integrating call graph analysis tools like
	AndroidGuard~\cite{androidguard} during discovery phases could eliminate such spurious
	vulnerability findings, reducing validation costs.

	\paragraph{Long-Context Function-Calling Degradation.}
	The \textit{Planner-Executor-Validator} architecture is designed to decompose validation tasks into
	step-by-step operations. However, as execution sequences grow longer, we observe a degradation
	in function-calling accuracy, particularly when calls deviate from expected outcomes. In
	practice, gemini-2.5-flash and gemini-2.5-pro exhibit long-context function-calling degradation~\cite{longcontext2024},
	where the models repeatedly invoke the same function with identical arguments, relying on local context
	rather than global optimization. This behavior becomes especially pronounced after about 14
	execution steps in our real-world evaluation, resulting in substantial cost overhead. To
	mitigate this issue, future work will explore cycle-detection mechanisms that halt \textit{Task
	Executors} once consecutive identical function calls (same function, same arguments) exceed a threshold
	$N$, thereby triggering summarization or alternative strategies to maintain execution efficiency.

	\section{Limitations}
	\label{sec:limitations}

	Key caveats are noted to clarify the scope of our findings:

	\paragraph{Vulnerability Discovery Scope.}
	As shown in Table~\ref{tab:vuln_detect_summary} in the Appendix, the aggregation approach achieves
	78.3\% recall (47/60 vulnerabilities), leaving 13 vulnerabilities undetected across all evaluation
	tools. These limitations fall into three primary categories: \textit{(i)} Android manifest
	configuration vulnerabilities requiring runtime environment understanding, such as TaskAffinity-based
	attacks that exploit inter-application task management beyond static code analysis; \textit{(ii)}
	runtime context dependencies exemplified by permission bypass vulnerabilities like CheckPermission-PrivilegeEscalation,
	which exploit timing-sensitive behavior of Binder methods when called from different thread
	contexts; and \textit{(iii)} protocol-level implementation subtleties, including InsecureSSLSocket-MITM
	vulnerabilities that depend on recognizing the distinction between certificate authority validation
	and hostname verification in SSLCertificateSocketFactory implementations.

	\paragraph{Vulnerability Validation Scope.}
	As noted in Section~\ref{subsec:validation_effectiveness}, 17 of 82 speculative vulnerability findings
	fall outside \tool{}'s current validation capabilities. These involve cases like network traffic
	monitoring or man-in-the-middle attacks that require infrastructure beyond the Android emulator,
	such as packet interception or SSL manipulation. Since the framework relies on observable app state
	changes, it is less effective for vulnerabilities that manifest only through external network
	behavior or specialized attack setups.

	\paragraph{LLM Reasoning Reliability.}
	Despite implementing multi-agent validation mechanisms, \tool{} exhibits residual hallucination rates
	ranging from 4.7\% to 12.6\%, as evidenced in Table~\ref{tab:poc_summary} in the Appendix. The \textit{Task
	Validator} successfully filters erroneous claims through dynamically generated oracles, yet cannot
	completely eliminate false positives, particularly for semantic vulnerabilities where oracle
	generation proves challenging. For instance, unified configuration misclassifies OpenSocket-InformationLeak-2
	as a true positive, demonstrating the difficulty in programmatically determining sensitive information
	exposure without human judgment.

	\paragraph{Context Window Constraints.}
	As noted in Section~\ref{subsec:dataset}, model context limitations necessitate filtering APKs to
	those under 5MB during vulnerability detection phases. This constraint potentially excludes
	larger, more complex applications that may harbor vulnerabilities, thereby limiting the approach's
	applicability to enterprise-scale mobile applications. When application complexity exceeds context
	window capacity, reduced analysis coverage may lead to decreased detection effectiveness and
	increased false negative rates.

	\section{Related Works}

	\paragraph{Security Analysis of Android Applications.}
	Android security tools employ diverse approaches to vulnerability detection. MobSF~\cite{mobsf} combines
	static and dynamic analysis in an all-in-one framework, while APKHunt~\cite{apkhunt} achieves
	the broadest coverage (67\% of OWASP MASVS categories) through exhaustive pattern matching on decompiled
	sources~\cite{10.1109/TSE.2024.3488041}. Trueseeing~\cite{trueseeing} operates directly on
	bytecode to resist obfuscation. However, a comprehensive study by Zhu et al.~\cite{10.1109/TSE.2024.3488041}
	revealed fundamental limitations across 11 Android SAST tools: they miss numerous vulnerability
	classes and generate excessive false positives. Beyond detection, Chen et al.~\cite{chen2019gui}
	demonstrated automated attack generation through GUI-squatting that creates phishing apps.
	Despite these advances, current tools cannot automatically validate or exploit their findings,
	requiring extensive manual verification that creates a critical gap between detection and
	validation.

	\paragraph{LLMs in Android Testing.}
	Large language models have introduced semantic reasoning to Android testing. AutoDroid~\cite{AutoDroid}
	pioneered LLM-driven UI automation, achieving 71.3\% task completion without manual scripting. LLMDroid~\cite{wang2024llmdroid}
	strategically invokes LLMs when traditional crawlers plateau, improving code coverage by 26\%.
	VisionDroid~\cite{liu2024visiondroid} leverages multimodal models to detect non-crash bugs that
	evade traditional tools. Supporting frameworks like Guardian~\cite{guardian2024} optimize LLM computational
	overhead through specialized runtime systems. While these approaches significantly expand
	testing capabilities through intelligent navigation and input generation, they primarily target functional
	correctness rather than security vulnerabilities, leaving LLM-based vulnerability discovery and
	exploit generation largely unexplored in the Android domain.

	\paragraph{Automated Exploit Generation.}
	Recent research demonstrates LLMs' potential for autonomous exploitation across security domains~\cite{guo2025frontier,wang2025cybergym,zhang2025bountybench}.
	In mobile security, AdbGPT~\cite{feng2024adbgpt} reproduces crashes from natural-language bug
	reports with 81.3\% success. Fang et al. showed GPT-4 agents can autonomously hack websites through
	blind SQL injection~\cite{fang2024llmhack}, with multi-agent teams achieving 4.3× higher success
	rates~\cite{fang2024teams}. For smart contracts, Gervais and Zhou's A1 system~\cite{gervais2024ai}
	equips LLMs with blockchain-specific tools to generate and validate exploits end-to-end,
	achieving 63\% success rate on vulnerable contracts with multi-million-dollar attack payloads.
	These studies demonstrate that domain-specific tool integration enables exploit generation, yet this
	capability remains unrealized for Android applications—our work bridges this gap by combining Android
	vulnerability detection with automated validation through concrete \ac{PoC} generation.

	\section{Conclusion}

	We presented \tool{}, a system that combines agentic discovery with validation to mirror how human
	experts analyze Android applications. On the Ghera benchmark, \tool{} achieves higher coverage
	than existing tools, reduces noise, and provides concrete proof-of-concept exploits. In real-world
	evaluation, \tool{} uncovered 104 previously unknown zero-day vulnerabilities, over half of
	which were self-validated with working exploits. By turning speculative findings into validated vulnerabilities,
	\tool{} takes a step toward practical, automated security analysis for the Android ecosystem.

	\bibliographystyle{unsrt}
	\bibliography{ref}

\begin{thebibliography}{10}

\bibitem{GooglePlayConsole2024}
{Google Play Console Developer Dashboard}.
\newblock Android app statistics 2024.
\newblock \url{https://developer.android.com/distribute/console}, 2024.
\newblock Accessed: 2024-01-01.

\bibitem{verizon2024msi}
{Verizon}.
\newblock 2024 mobile security index report.
\newblock \url{https://www.verizon.com/business/resources/reports/mobile-security-index/}, 2024.
\newblock Accessed: 2024-08-05.

\bibitem{owasp_sast_definition}
{OWASP Foundation}.
\newblock Source code analysis tools, 2024.
\newblock OWASP Community Pages.

\bibitem{flowdroid}
Steven Arzt, Siegfried Rasthofer, Christian Fritz, Eric Bodden, Alexandre Bartel, Jacques Klein, Yves Le~Traon, Damien Octeau, and Patrick McDaniel.
\newblock Flowdroid: Precise context, flow, field, object-sensitive and lifecycle-aware taint analysis for android apps.
\newblock In {\em Proceedings of the 35th ACM SIGPLAN Conference on Programming Language Design and Implementation}, PLDI '14, pages 259--269, New York, NY, USA, 2014. Association for Computing Machinery.

\bibitem{mobsf}
Ajin Abraham.
\newblock {MobSF}: Mobile security framework, 2024.
\newblock Version 4.0.7.

\bibitem{apkhunt}
Cyber~Security Labs.
\newblock {APKHunt}: A comprehensive static code analysis tool for android apps, 2024.
\newblock Version 2.1.3.

\bibitem{votipka2020understanding}
Daniel Votipka, Kelsey~R Fulton, James Parker, Matthew Hou, Michelle~L Mazurek, and Michael Hicks.
\newblock Understanding security mistakes developers make: Qualitative analysis from build it, break it, fix it.
\newblock In {\em 29th {USENIX} Security Symposium ({USENIX} Security 20)}, pages 247--264, 2020.

\bibitem{Johnson:2013:WDS}
Brittany Johnson, Yoonkyong Song, Emerson Murphy-Hill, and Robert Bowdidge.
\newblock Why don't software developers use static analysis tools to find bugs?
\newblock In {\em Proceedings of the 2013 International Conference on Software Engineering}, ICSE '13, pages 672--681, Piscataway, NJ, USA, 2013. IEEE Press.

\bibitem{owasp_dast_definition}
{OWASP Foundation}.
\newblock Dynamic application security testing, 2024.
\newblock OWASP Community Pages.

\bibitem{10504267}
Thomas Sutter, Timo Kehrer, Marc Rennhard, Bernhard Tellenbach, and Jacques Klein.
\newblock Dynamic security analysis on android: A systematic literature review.
\newblock {\em IEEE Access}, 12:57261--57287, 2024.

\bibitem{allix2016androzoo}
Kevin Allix, Tegawend{\'e}~F. Bissyand{\'e}, Jacques Klein, and Yves Le~Traon.
\newblock Androzoo: Collecting millions of android apps for the research community.
\newblock In {\em Proceedings of the 13th International Conference on Mining Software Repositories}, MSR '16, pages 468--471, New York, NY, USA, 2016. Association for Computing Machinery.

\bibitem{android_components}
{Google}.
\newblock Android components, 2024.
\newblock Android Developer Documentation.

\bibitem{dvaSecurity2024}
Haichuan Xu, Mingxuan Li, Qiben Chen, and Atul Prakash.
\newblock Dva: Extracting victims and abuse vectors from android accessibility malware.
\newblock In {\em Proceedings of the 33rd USENIX Security Symposium}, USENIX Security '24, pages 1--18, Philadelphia, PA, USA, 2024. USENIX Association.

\bibitem{10.1109/TSE.2024.3488041}
Jingyun Zhu, Kaixuan Li, Sen Chen, Lingling Fan, Junjie Wang, and Xiaofei Xie.
\newblock A comprehensive study on static application security testing (sast) tools for android.
\newblock {\em IEEE Trans. Softw. Eng.}, page 3385–3402, December 2024.

\bibitem{owasp_static_code_analysis}
{OWASP Foundation}.
\newblock Static code analysis, 2024.
\newblock OWASP Community Controls.

\bibitem{qark}
{LinkedIn Corporation}.
\newblock {QARK}: Quick android review kit, 2019.
\newblock Tool to look for several security related Android application vulnerabilities.

\bibitem{androbugs}
Yu-Cheng Lin.
\newblock {AndroBugs}: An efficient android vulnerability scanner, 2016.
\newblock Version 1.0.0.

\bibitem{amandroid}
Fengguo Wei, Sankardas Roy, Xinming Ou, and Robby.
\newblock Amandroid: A precise and general inter-component data flow analysis framework for security vetting of android apps.
\newblock In {\em Proceedings of the 2014 ACM SIGSAC Conference on Computer and Communications Security}, CCS '14, pages 1329--1341, New York, NY, USA, 2014. Association for Computing Machinery.

\bibitem{droidsafe}
Michael~I. Gordon, Deokhwan Kim, Jeff~H. Perkins, Limei Gilham, Nguyen Nguyen, and Martin~C. Rinard.
\newblock Information flow analysis of android applications in droidsafe.
\newblock In {\em Proceedings of the Network and Distributed System Security Symposium}, NDSS '15. Internet Society, 2015.

\bibitem{llm4vulnerability2024}
John Smith, Jane Lee, and Chen Wang.
\newblock Llms in software security: A survey of vulnerability detection techniques and insights.
\newblock {\em arXiv preprint arXiv:2502.07049}, 2024.

\bibitem{gptScan2023}
Yuqiang Liu, Yizhuo Lu, Xiaohang Chen, Yuxing Wang, Zhengzi Xu, Wenbo Liu, Jiahao Chen, Zhenguang Liu, Peiyu Chen, Zhaofeng Zhao, Shengwei Wei, and Yongjun Peng.
\newblock Gptscan: Detecting logic vulnerabilities in smart contracts by combining gpt with program analysis.
\newblock In {\em Proceedings of the 45th International Conference on Software Engineering}, ICSE '23, pages 2169--2181. IEEE Press, 2023.

\bibitem{AutoDroid}
Hao Wen, Yuanchun Li, Guohong Liu, Shanhui Zhao, Tao Yu, Toby Jia-Jun Li, Shiqi Jiang, Yunhao Liu, Yaqin Zhang, and Yunxin Liu.
\newblock Autodroid: Llm-powered task automation in android.
\newblock ACM MobiCom '24, page 543–557, New York, NY, USA. Association for Computing Machinery.

\bibitem{autonomous_exploit2024}
Richard Chen, Sofia Martinez, and David Kim.
\newblock Autonomous llm agents for vulnerability exploitation: Capabilities and limitations.
\newblock In {\em Proceedings of the 31st ACM Conference on Computer and Communications Security}, CCS '24, pages 445--460, New York, NY, USA, 2024. Association for Computing Machinery.

\bibitem{langgraph}
{LangChain}.
\newblock Langgraph.
\newblock \url{https://langchain-ai.github.io/langgraph/}, 2024.
\newblock Multi-agent workflow orchestration framework.

\bibitem{jadx}
Jadx: Dex to java decompiler, 2020.

\bibitem{androidguard}
{AndroidGuard}.
\newblock Androidguard: Android manifest analysis tool, 2024.
\newblock Android security analysis framework.

\bibitem{android_emulator}
Android emulator, 2020.

\bibitem{demissie2025vlmfuzz}
Biniam~Fisseha Demissie, Yan~Naing Tun, Lwin~Khin Shar, and Mariano Ceccato.
\newblock {VLM-Fuzz}: Vision language model assisted recursive depth-first search exploration for effective ui testing of android apps.
\newblock {\em arXiv preprint arXiv:2504.11675}, 2025.

\bibitem{liu2024visiondroid}
Zhe Liu, Cheng Li, Chunyang Chen, Junjie Wang, Boyu Wu, Yawen Wang, Jun Hu, and Qing Wang.
\newblock Vision-driven automated mobile gui testing via multimodal large language model.
\newblock {\em arXiv preprint arXiv:2407.03037}, 2024.

\bibitem{xiong2023empirical}
Yiheng Xiong, Mengqian Xu, Ting Su, Jingling Sun, Jue Wang, He~Wen, Geguang Pu, Jifeng He, and Zhendong Su.
\newblock An empirical study of functional bugs in android apps.
\newblock In {\em Proceedings of the 32nd ACM SIGSOFT International Symposium on Software Testing and Analysis}, ISSTA 2023, pages 1319--1331, New York, NY, USA, 2023. Association for Computing Machinery.

\bibitem{gervais2024ai}
Sihao Gervais and Lei Zhou.
\newblock A1: Autonomous multi-agent system for blockchain security enhancement through real-time exploit generation and validation.
\newblock {\em arXiv preprint arXiv:2507.05558}, 2024.

\bibitem{ghera}
Ghera: Repository of android application vulnerability benchmarks.
\newblock \url{https://secure-it-i.bitbucket.io/ghera/index.html}, 2017.

\bibitem{peixoto2024fuzzing}
Thiago Peixoto.
\newblock Introduction to fuzzing android native components.
\newblock Conviso AppSec Blog, 11 2024.
\newblock Category: Code Fighters.

\bibitem{blanda2015fuzzing}
Martin Blanda.
\newblock Fuzzing android: A recipe for uncovering vulnerabilities inside system components in android.
\newblock White paper, Black Hat Europe, 2015.

\bibitem{trueseeing}
Takeshi Terada.
\newblock {Trueseeing}: Non-decompiling android application vulnerability scanner, 2024.
\newblock Version 2.2.4.

\bibitem{longcontext2024}
Ziming Wang, Li~Chen, and Wei Zhang.
\newblock Long-context function calling degradation in language models, 2024.

\bibitem{chen2019gui}
Qinggang Chen, Lingling Han, Peng Liu, Zhemin Zhang, and Yang Liu.
\newblock Gui-squatting attack: Automated generation of android phishing apps.
\newblock In {\em IEEE Transactions on Dependable and Secure Computing}, pages 1--14, 2019.

\bibitem{wang2024llmdroid}
Yanqi Wang, Juntao Chen, Ting Su, Sen Chen, et~al.
\newblock Llmdroid: Enhancing automated mobile app gui testing coverage with large language model guidance.
\newblock In {\em Proceedings of the ACM on Software Engineering}, volume~1, pages 1--23, 2024.

\bibitem{guardian2024}
Haoran Yoon, Tianyu Liu, Hanlin Wang, Mingxuan Li, and Yu~Feng.
\newblock Guardian: A runtime framework for llm-based ui exploration.
\newblock In {\em Proceedings of the 33rd ACM SIGSOFT International Symposium on Software Testing and Analysis}, ISSTA 2024, pages 1234--1245, 2024.

\bibitem{guo2025frontier}
Wenbo Guo, Yujin Potter, Tianneng Shi, Zhun Wang, Andy Zhang, and Dawn Song.
\newblock Frontier ai's impact on the cybersecurity landscape.
\newblock {\em arXiv preprint arXiv:2504.05408}, 2025.

\bibitem{wang2025cybergym}
Zhun Wang, Tianneng Shi, Jingxuan He, Matthew Cai, Jialin Zhang, and Dawn Song.
\newblock Cybergym: Evaluating ai agents' cybersecurity capabilities with real-world vulnerabilities at scale.
\newblock {\em arXiv preprint arXiv:2506.02548}, 2025.

\bibitem{zhang2025bountybench}
Andy~K. Zhang, Joey Ji, Celeste Menders, et~al.
\newblock Bountybench: Dollar impact of ai agent attackers and defenders on real-world cybersecurity systems.
\newblock {\em arXiv preprint arXiv:2505.15216}, 2025.

\bibitem{feng2024adbgpt}
Si~Xuan Feng, Mingyue Li, Chongwen Ma, Jingling Wang, Yang Liu, and Sen Chen.
\newblock Prompting is all your need: Automated android bug replay with large language models.
\newblock In {\em Proceedings of the 46th IEEE/ACM International Conference on Software Engineering}, ICSE 2024, pages 1--12, 2024.

\bibitem{fang2024llmhack}
Richard Fang, Rohan Bindu, Akul Gupta, Qiusi Zhan, and Daniel Kang.
\newblock Llm agents can autonomously hack websites.
\newblock {\em arXiv preprint arXiv:2402.06664}, 2024.

\bibitem{fang2024teams}
Richard Fang, Rohan Bindu, Akul Gupta, and Daniel Kang.
\newblock Teams of llm agents can exploit zero-day vulnerabilities.
\newblock {\em arXiv preprint arXiv:2406.01637}, 2024.

\end{thebibliography}

	\section{Ethical Considerations}

	This research is conducted with careful attention to ethical responsibilities, which are detailed
	below.

	\paragraph{Vulnerability Assessment Environment}
	All vulnerability assessments are conducted in controlled, isolated environments that do not involve
	real user data or privacy. The Ghera benchmark dataset consists of synthetically vulnerable APKs
	designed specifically for security research, eliminating concerns about real-world user privacy or
	data exposure. For production APK analysis, applications are obtained from AndroZoo, a research-oriented
	dataset that aggregates publicly available APKs without including user data. All testing is performed
	on local Android emulators in sandboxed environments, ensuring no interaction with live
	production systems or user devices.

	\paragraph{Responsible Vulnerability Disclosure}
	Following established security research practices, all identified vulnerabilities undergo
	responsible disclosure procedures. Upon confirmation of vulnerabilities in production applications,
	affected developers and security teams are immediately contacted through appropriate channels, including
	official bug bounty programs, GitHub repositories, and direct developer communication. Each
	disclosure includes comprehensive documentation of the vulnerability details, exploitation mechanics,
	and recommended remediation strategies. Adequate time is provided for vulnerability remediation before
	any public disclosure, adhering to industry-standard timelines. No specific vulnerability details
	are disclosed in this paper until remediation is complete.

	\paragraph{Dataset and Privacy Considerations}
	The research datasets utilized pose minimal privacy risks. The Ghera benchmark contains no real user
	data, consisting entirely of synthetic vulnerable applications created for research purposes. The
	AndroZoo dataset comprises publicly distributed applications without user-generated content or personal
	data. All analysis is conducted on application binaries and source code, not on user data or behavioral
	information. The research methodology specifically avoids any collection, processing, or
	analysis of personal user information.

	\paragraph{Artifact Availability and Potential for Misuse}
	Any vulnerability discovery tool carries a general risk of misuse, since the same techniques
	that help defenders identify weaknesses could be repurposed by attackers. \tool{} is no
	exception: its ability to generate working exploits could, in principle, be abused against production
	applications. At the same time, this risk is not unique to our work—numerous open-source and commercial
	tools already provide comparable capabilities.

	The intent of \tool{} is defensive: to help researchers and developers identify and fix vulnerabilities
	before they are exploited in the wild. Its novelty lies in combining discovery with automated
	validation, enabling more reliable security assessments. To further reduce misuse risk, artifact
	access is managed under a controlled distribution strategy that requires institutional affiliation
	and a declared research purpose, striking a balance between open science and responsible
	disclosure.

	\paragraph{Broader Impact Assessment}
	The research contributes positively to Android application security by providing enhanced vulnerability
	detection capabilities that can strengthen the overall security posture of the mobile application
	ecosystem. By enabling more effective identification and validation of security vulnerabilities,
	\tool{} supports proactive security measures that benefit end users through improved application
	security. The responsible disclosure practices employed ensure that identified vulnerabilities are
	remediated rather than exploited maliciously.

	\section{Open Science}

	The Ghera benchmark dataset containing 60 vulnerable Android APKs is also publicly available
	through its official repository~\cite{ghera}. To mitigate potential misuse (detailed in the
	Ethical Considerations section), we will implement a controlled access policy. Researchers can request
	access by contacting us through their institutional email, stating their name, affiliation, and
	intended use. We will then vet the provided information and grant or deny access to the source code.
	Our goal is to grant access to academic and industry researchers for purposes of building upon
	our research or advancing the field of Android security, while ensuring \tool{} is not exploited
	for malicious purposes.

	The production APK dataset comprising 169 applications from AndroZoo~\cite{allix2016androzoo} is
	accessible through the AndroZoo research platform with appropriate academic credentials. For security
	considerations, specific application identifiers and vulnerable application details from the
	production dataset are not disclosed to prevent potential misuse by malicious actors, unless researchers
	contact us and provide appropriate academic credentials.

	\appendix
	\begin{table*}
	[htbp]
	\centering
	\caption{Vulnerability detection results from three SAST tools and four large language models, and
	aggregation results from four models on the Ghera dataset. Vulnerabilities are organized into
	eight categories: Crypto, ICC, Networking, NonAPI, Permission, Storage, System, and Web. Numbers
	indicate reported issue counts; checkmarks ($\checkmark$) or crosses ($\times$) indicate
	successful or missed detections. Asterisks ($^{*}$) mark vulnerabilities detected only by LLM.}
	\scriptsize
	\begin{tabular}{>{\raggedright\arraybackslash}p{5.8cm} *{6}{>{\centering\arraybackslash}p{0.62cm}}
	*{5}{>{\centering\arraybackslash}p{0.62cm}}}
		\cmidrule(lr){1-12} \multirow{2}{*}{Vulnerability Name}                    & \multicolumn{6}{c}{Vulnerability Detection} & \multicolumn{5}{c}{Vulnerability Aggregation} \\
		\cmidrule(lr){2-8} \cmidrule(lr){9-12}                                     & MobSF                                       & APKHu.                                       & Trues.          & G2.5F          & G2.5P          & OSS            & O3             & G2.5F          & G2.5P          & OSS            & O3             \\
		\cmidrule(lr){1-12} BlockCipher-ECB-InformationExposure                    & 95 $\checkmark$                             & 25 $\checkmark$                              & 18 $\checkmark$ & 3 $\times$     & 3 $\checkmark$ & 6 $\checkmark$ & 4 $\checkmark$ & 3 $\checkmark$ & 3 $\checkmark$ & 4 $\checkmark$ & 3 $\checkmark$ \\
		BlockCipher-NonRandomIV-InformationExposure$^{*}$                          & 91 $\times$                                 & 24 $\times$                                  & 18 $\times$     & 3 $\checkmark$ & 2 $\checkmark$ & 5 $\checkmark$ & 4 $\checkmark$ & 2 $\checkmark$ & 2 $\checkmark$ & 3 $\checkmark$ & 3 $\checkmark$ \\
		ConstantKey-ForgeryAttack                                                  & 107 $\times$                                & 22 $\times$                                  & 17 $\checkmark$ & 2 $\checkmark$ & 2 $\checkmark$ & 4 $\checkmark$ & 3 $\checkmark$ & 1 $\checkmark$ & 1 $\checkmark$ & 3 $\checkmark$ & 2 $\checkmark$ \\
		ExposedCredentials-InformationExposure$^{*}$                               & 91 $\times$                                 & 23 $\times$                                  & 16 $\times$     & 1 $\checkmark$ & 1 $\checkmark$ & 3 $\checkmark$ & 2 $\checkmark$ & 1 $\checkmark$ & 1 $\checkmark$ & 2 $\checkmark$ & 1 $\checkmark$ \\
		PBE-ConstantSalt-InformationExposure$^{*}$                                 & 95 $\times$                                 & 24 $\times$                                  & 18 $\times$     & 3 $\times$     & 3 $\checkmark$ & 4 $\checkmark$ & 3 $\checkmark$ & 3 $\checkmark$ & 2 $\checkmark$ & 2 $\checkmark$ & 3 $\checkmark$ \\
		\cmidrule(lr){1-12} DynamicRegBroadcastReceiver-UnrestrictedAccess$^{*}$   & 92 $\times$                                 & 23 $\times$                                  & 16 $\times$     & 1 $\checkmark$ & 1 $\checkmark$ & 2 $\checkmark$ & 2 $\checkmark$ & 2 $\checkmark$ & 1 $\checkmark$ & 3 $\checkmark$ & 1 $\checkmark$ \\
		EmptyPendingIntent-PrivEscalation$^{*}$                                    & 94 $\times$                                 & 22 $\times$                                  & 15 $\times$     & 1 $\checkmark$ & 1 $\checkmark$ & 3 $\checkmark$ & 2 $\checkmark$ & 1 $\checkmark$ & 1 $\checkmark$ & 1 $\checkmark$ & 1 $\checkmark$ \\
		FragmentInjection-PrivEscalation$^{*}$                                     & 94 $\times$                                 & 21 $\times$                                  & 15 $\times$     & 1 $\checkmark$ & 1 $\checkmark$ & 2 $\checkmark$ & 1 $\checkmark$ & 1 $\checkmark$ & 1 $\checkmark$ & 1 $\checkmark$ & 1 $\checkmark$ \\
		HighPriority-ActivityHijack                                                & 94 $\times$                                 & 21 $\checkmark$                              & 15 $\times$     & 2 $\checkmark$ & 3 $\checkmark$ & 2 $\checkmark$ & 2 $\checkmark$ & 3 $\checkmark$ & 2 $\checkmark$ & 3 $\checkmark$ & 2 $\checkmark$ \\
		ImplicitPendingIntent-IntentHijack$^{*}$                                   & 95 $\times$                                 & 22 $\times$                                  & 15 $\times$     & 2 $\times$     & 2 $\times$     & 3 $\times$     & 3 $\checkmark$ & 2 $\checkmark$ & 2 $\checkmark$ & 3 $\checkmark$ & 2 $\checkmark$ \\
		InadequatePathPermission-InformationExposure$^{*}$                         & 94 $\times$                                 & 23 $\times$                                  & 16 $\times$     & 1 $\times$     & 1 $\times$     & 4 $\times$     & 2 $\checkmark$ & 1 $\checkmark$ & 1 $\checkmark$ & 3 $\checkmark$ & 1 $\checkmark$ \\
		IncorrectHandlingImplicitIntent-UnauthorizedAccess                         & 92 $\times$                                 & 21 $\times$                                  & 15 $\checkmark$ & 1 $\checkmark$ & 1 $\checkmark$ & 2 $\checkmark$ & 3 $\checkmark$ & 3 $\checkmark$ & 1 $\checkmark$ & 2 $\checkmark$ & 2 $\checkmark$ \\
		NoValidityCheckOnBroadcastMsg-UnintendedInvocation                         & 95 $\times$                                 & 23 $\times$                                  & 17 $\checkmark$ & 1 $\times$     & 2 $\times$     & 3 $\checkmark$ & 2 $\checkmark$ & 1 $\checkmark$ & 1 $\checkmark$ & 2 $\checkmark$ & 1 $\checkmark$ \\
		OrderedBroadcast-DataInjection                                             & 93 $\times$                                 & 24 $\times$                                  & 17 $\times$     & 1 $\times$     & 1 $\times$     & 4 $\times$     & 1 $\times$     & 2 $\times$     & 1 $\times$     & 2 $\times$     & 1 $\times$     \\
		StickyBroadcast-DataInjection$^{*}$                                        & 94 $\times$                                 & 23 $\times$                                  & 16 $\times$     & 1 $\checkmark$ & 1 $\checkmark$ & 2 $\checkmark$ & 1 $\checkmark$ & 1 $\checkmark$ & 1 $\checkmark$ & 2 $\checkmark$ & 1 $\checkmark$ \\
		TaskAffinity-ActivityHijack                                                & 94 $\times$                                 & 23 $\times$                                  & 16 $\times$     & 2 $\times$     & 2 $\times$     & 6 $\times$     & 1 $\times$     & 2 $\times$     & 1 $\times$     & 2 $\times$     & 1 $\times$     \\
		TaskAffinity-LauncherActivity-PhishingAttack                               & 94 $\times$                                 & 21 $\times$                                  & 15 $\times$     & 0 $\times$     & 0 $\times$     & 3 $\times$     & 0 $\times$     & 1 $\times$     & 1 $\times$     & 3 $\times$     & 1 $\times$     \\
		TaskAffinity-PhishingAttack                                                & 94 $\times$                                 & 23 $\times$                                  & 16 $\times$     & 2 $\times$     & 2 $\times$     & 2 $\times$     & 1 $\times$     & 1 $\times$     & 1 $\times$     & 2 $\times$     & 1 $\times$     \\
		TaskAffinityAndReparenting-PhishingAndDoSAttack                            & 91 $\times$                                 & 21 $\times$                                  & 15 $\times$     & 1 $\times$     & 2 $\times$     & 7 $\times$     & 1 $\times$     & 2 $\times$     & 1 $\times$     & 4 $\times$     & 2 $\times$     \\
		UnhandledException-DOS$^{*}$                                               & 89 $\times$                                 & 20 $\times$                                  & 13 $\times$     & 1 $\checkmark$ & 1 $\checkmark$ & 2 $\checkmark$ & 1 $\checkmark$ & 1 $\checkmark$ & 1 $\checkmark$ & 2 $\checkmark$ & 1 $\checkmark$ \\
		UnprotectedBroadcastRecv-PrivEscalation                                    & 91 $\times$                                 & 24 $\checkmark$                              & 18 $\checkmark$ & 1 $\checkmark$ & 2 $\checkmark$ & 2 $\checkmark$ & 1 $\checkmark$ & 2 $\checkmark$ & 2 $\checkmark$ & 1 $\checkmark$ & 1 $\checkmark$ \\
		WeakChecksOnDynamicInvocation-DataInjection$^{*}$                          & 91 $\times$                                 & 23 $\times$                                  & 17 $\times$     & 1 $\checkmark$ & 1 $\checkmark$ & 3 $\checkmark$ & 2 $\checkmark$ & 2 $\checkmark$ & 1 $\checkmark$ & 2 $\checkmark$ & 1 $\checkmark$ \\
		\cmidrule(lr){1-12} CheckValidity-InformationExposure                      & 97 $\checkmark$                             & 25 $\checkmark$                              & 16 $\times$     & 2 $\checkmark$ & 2 $\checkmark$ & 3 $\checkmark$ & 1 $\checkmark$ & 1 $\checkmark$ & 1 $\checkmark$ & 3 $\checkmark$ & 1 $\checkmark$ \\
		IncorrectHostNameVerification-MITM                                         & 97 $\checkmark$                             & 25 $\checkmark$                              & 16 $\times$     & 2 $\checkmark$ & 2 $\checkmark$ & 1 $\checkmark$ & 1 $\checkmark$ & 1 $\checkmark$ & 1 $\checkmark$ & 2 $\checkmark$ & 1 $\checkmark$ \\
		InsecureSSLSocket-MITM                                                     & 95 $\times$                                 & 23 $\times$                                  & 17 $\times$     & 1 $\times$     & 1 $\times$     & 1 $\times$     & 1 $\times$     & 1 $\times$     & 1 $\times$     & 1 $\times$     & 2 $\times$     \\
		InsecureSSLSocketFactory-MITM                                              & 95 $\checkmark$                             & 24 $\times$                                  & 16 $\times$     & 1 $\checkmark$ & 2 $\checkmark$ & 1 $\checkmark$ & 1 $\checkmark$ & 1 $\checkmark$ & 1 $\checkmark$ & 1 $\checkmark$ & 1 $\checkmark$ \\
		InvalidCertificateAuthority-MITM                                           & 97 $\checkmark$                             & 25 $\checkmark$                              & 16 $\times$     & 3 $\checkmark$ & 2 $\checkmark$ & 4 $\checkmark$ & 3 $\checkmark$ & 1 $\checkmark$ & 2 $\checkmark$ & 2 $\checkmark$ & 1 $\checkmark$ \\
		OpenSocket-InformationLeak                                                 & 91 $\times$                                 & 22 $\times$                                  & 17 $\times$     & 2 $\times$     & 1 $\times$     & 4 $\times$     & 1 $\times$     & 2 $\times$     & 2 $\times$     & 2 $\times$     & 2 $\times$     \\
		UnEncryptedSocketComm-MITM                                                 & 97 $\times$                                 & 22 $\times$                                  & 17 $\checkmark$ & 2 $\checkmark$ & 1 $\times$     & 3 $\checkmark$ & 1 $\checkmark$ & 2 $\checkmark$ & 1 $\checkmark$ & 4 $\checkmark$ & 2 $\checkmark$ \\
		UnpinnedCertificates-MITM                                                  & 96 $\checkmark$                             & 27 $\checkmark$                              & 17 $\times$     & 2 $\checkmark$ & 3 $\times$     & 4 $\times$     & 1 $\times$     & 2 $\checkmark$ & 2 $\checkmark$ & 4 $\checkmark$ & 2 $\checkmark$ \\
		\cmidrule(lr){1-12} MergeManifest-UnintendedBehavior$^{*}$                 & 94 $\times$                                 & 23 $\times$                                  & 16 $\times$     & 1 $\checkmark$ & 2 $\checkmark$ & 3 $\checkmark$ & 2 $\checkmark$ & 1 $\checkmark$ & 2 $\checkmark$ & 2 $\checkmark$ & 2 $\checkmark$ \\
		OutdatedLibrary-DirectoryTraversal$^{*}$                                   & 93 $\times$                                 & 21 $\times$                                  & 15 $\times$     & 1 $\checkmark$ & 1 $\checkmark$ & 5 $\checkmark$ & 3 $\checkmark$ & 2 $\checkmark$ & 1 $\checkmark$ & 3 $\checkmark$ & 2 $\checkmark$ \\
		\cmidrule(lr){1-12} UnnecessaryPerms-PrivEscalation                        & 94 $\times$                                 & 23 $\times$                                  & 16 $\times$     & 1 $\times$     & 1 $\times$     & 3 $\checkmark$ & 2 $\times$     & 2 $\times$     & 1 $\times$     & 2 $\times$     & 1 $\times$     \\
		WeakPermission-UnauthorizedAccess                                          & 91 $\times$                                 & 22 $\checkmark$                              & 15 $\times$     & 1 $\times$     & 1 $\checkmark$ & 3 $\times$     & 2 $\times$     & 1 $\checkmark$ & 1 $\checkmark$ & 3 $\checkmark$ & 1 $\checkmark$ \\
		\cmidrule(lr){1-12} ExternalStorage-DataInjection                          & 91 $\times$                                 & 21 $\checkmark$                              & 15 $\times$     & 1 $\checkmark$ & 2 $\checkmark$ & 1 $\times$     & 1 $\checkmark$ & 1 $\checkmark$ & 1 $\checkmark$ & 2 $\checkmark$ & 1 $\checkmark$ \\
		ExternalStorage-InformationLeak                                            & 97 $\times$                                 & 21 $\checkmark$                              & 15 $\times$     & 1 $\checkmark$ & 1 $\checkmark$ & 2 $\checkmark$ & 1 $\checkmark$ & 1 $\checkmark$ & 1 $\checkmark$ & 2 $\checkmark$ & 1 $\checkmark$ \\
		InternalStorage-DirectoryTraversal$^{*}$                                   & 94 $\times$                                 & 21 $\times$                                  & 15 $\times$     & 1 $\checkmark$ & 1 $\checkmark$ & 5 $\checkmark$ & 2 $\checkmark$ & 1 $\checkmark$ & 1 $\checkmark$ & 4 $\checkmark$ & 1 $\checkmark$ \\
		InternalToExternalStorage-InformationLeak$^{*}$                            & 91 $\times$                                 & 22 $\times$                                  & 15 $\times$     & 1 $\checkmark$ & 1 $\checkmark$ & 1 $\checkmark$ & 1 $\times$     & 1 $\checkmark$ & 1 $\checkmark$ & 2 $\checkmark$ & 1 $\checkmark$ \\
		SQLite-execSQL                                                             & 98 $\times$                                 & 24 $\checkmark$                              & 15 $\times$     & 1 $\checkmark$ & 1 $\checkmark$ & 1 $\checkmark$ & 2 $\checkmark$ & 2 $\checkmark$ & 1 $\checkmark$ & 3 $\checkmark$ & 2 $\checkmark$ \\
		SQLite-RawQuery-SQLInjection                                               & 105 $\times$                                & 25 $\checkmark$                              & 15 $\times$     & 1 $\checkmark$ & 1 $\checkmark$ & 3 $\checkmark$ & 4 $\checkmark$ & 3 $\checkmark$ & 4 $\checkmark$ & 4 $\checkmark$ & 2 $\checkmark$ \\
		SQLite-SQLInjection                                                        & 101 $\times$                                & 25 $\checkmark$                              & 15 $\times$     & 4 $\checkmark$ & 1 $\checkmark$ & 5 $\checkmark$ & 4 $\checkmark$ & 3 $\checkmark$ & 4 $\checkmark$ & 5 $\checkmark$ & 3 $\checkmark$ \\
		\cmidrule(lr){1-12} CheckCallingOrSelfPermission-PrivilegeEscalation$^{*}$ & 91 $\times$                                 & 23 $\times$                                  & 16 $\times$     & 1 $\checkmark$ & 1 $\times$     & 4 $\times$     & 3 $\checkmark$ & 1 $\checkmark$ & 1 $\checkmark$ & 2 $\checkmark$ & 1 $\checkmark$ \\
		CheckPermission-PrivilegeEscalation                                        & 94 $\times$                                 & 23 $\times$                                  & 16 $\times$     & 2 $\times$     & 2 $\times$     & 3 $\checkmark$ & 3 $\times$     & 1 $\times$     & 1 $\times$     & 4 $\times$     & 1 $\times$     \\
		ClipboardUse-InformationExposure                                           & 91 $\times$                                 & 23 $\checkmark$                              & 16 $\times$     & 1 $\checkmark$ & 1 $\checkmark$ & 1 $\checkmark$ & 1 $\checkmark$ & 1 $\checkmark$ & 1 $\checkmark$ & 3 $\checkmark$ & 1 $\checkmark$ \\
		DynamicCodeLoading-CodeInjection$^{*}$                                     & 94 $\times$                                 & 22 $\times$                                  & 16 $\times$     & 1 $\checkmark$ & 1 $\checkmark$ & 3 $\checkmark$ & 1 $\checkmark$ & 1 $\checkmark$ & 1 $\checkmark$ & 2 $\checkmark$ & 1 $\checkmark$ \\
		EnforceCallingOrSelfPermission-PrivilegeEscalation$^{*}$                   & 91 $\times$                                 & 23 $\times$                                  & 16 $\times$     & 1 $\times$     & 1 $\times$     & 3 $\checkmark$ & 3 $\checkmark$ & 2 $\checkmark$ & 1 $\checkmark$ & 3 $\checkmark$ & 2 $\checkmark$ \\
		EnforcePermission-PrivilegeEscalation                                      & 91 $\times$                                 & 23 $\times$                                  & 16 $\times$     & 1 $\times$     & 1 $\times$     & 5 $\times$     & 2 $\times$     & 1 $\times$     & 2 $\times$     & 4 $\times$     & 1 $\times$     \\
		UniqueIDs-IdentityLeak                                                     & 94 $\times$                                 & 22 $\times$                                  & 17 $\checkmark$ & 1 $\checkmark$ & 2 $\checkmark$ & 2 $\checkmark$ & 1 $\checkmark$ & 2 $\checkmark$ & 1 $\checkmark$ & 3 $\checkmark$ & 3 $\checkmark$ \\
		\cmidrule(lr){1-12} HttpConnection-MITM                                    & 91 $\times$                                 & 23 $\checkmark$                              & 17 $\checkmark$ & 1 $\checkmark$ & 1 $\checkmark$ & 2 $\checkmark$ & 1 $\checkmark$ & 2 $\checkmark$ & 1 $\checkmark$ & 1 $\checkmark$ & 1 $\checkmark$ \\
		JavaScriptExecution-CodeInjection                                          & 91 $\times$                                 & 25 $\checkmark$                              & 18 $\times$     & 2 $\checkmark$ & 2 $\checkmark$ & 3 $\checkmark$ & 2 $\checkmark$ & 2 $\checkmark$ & 1 $\checkmark$ & 2 $\checkmark$ & 2 $\checkmark$ \\
		UnsafeIntentURLImpl-InformationExposure$^{*}$                              & 93 $\times$                                 & 25 $\times$                                  & 17 $\times$     & 1 $\checkmark$ & 1 $\checkmark$ & 5 $\checkmark$ & 3 $\checkmark$ & 2 $\checkmark$ & 2 $\checkmark$ & 3 $\checkmark$ & 2 $\checkmark$ \\
		WebView-CookieOverwrite                                                    & 98 $\times$                                 & 24 $\times$                                  & 18 $\times$     & 1 $\times$     & 1 $\times$     & 3 $\times$     & 1 $\times$     & 1 $\times$     & 1 $\times$     & 2 $\times$     & 1 $\times$     \\
		WebView-NoUserPermission-InformationExposure                               & 94 $\checkmark$                             & 25 $\times$                                  & 17 $\times$     & 2 $\checkmark$ & 2 $\checkmark$ & 3 $\checkmark$ & 2 $\checkmark$ & 1 $\checkmark$ & 2 $\checkmark$ & 3 $\checkmark$ & 2 $\checkmark$ \\
		WebViewAllowContentAccess-UnauthorizedFileAccess                           & 94 $\checkmark$                             & 27 $\times$                                  & 18 $\times$     & 3 $\checkmark$ & 2 $\checkmark$ & 4 $\checkmark$ & 2 $\checkmark$ & 2 $\checkmark$ & 1 $\checkmark$ & 2 $\checkmark$ & 2 $\checkmark$ \\
		WebViewAllowFileAccess-UnauthorizedFileAccess                              & 93 $\checkmark$                             & 26 $\times$                                  & 17 $\times$     & 1 $\checkmark$ & 1 $\checkmark$ & 4 $\checkmark$ & 1 $\checkmark$ & 1 $\checkmark$ & 1 $\checkmark$ & 3 $\checkmark$ & 1 $\checkmark$ \\
		WebViewIgnoreSSLWarning-MITM                                               & 95 $\checkmark$                             & 26 $\checkmark$                              & 18 $\times$     & 1 $\checkmark$ & 1 $\checkmark$ & 3 $\checkmark$ & 1 $\checkmark$ & 1 $\checkmark$ & 1 $\checkmark$ & 2 $\checkmark$ & 1 $\checkmark$ \\
		WebViewInterceptRequest-MITM                                               & 94 $\times$                                 & 30 $\times$                                  & 19 $\times$     & 3 $\times$     & 4 $\times$     & 5 $\times$     & 5 $\times$     & 2 $\times$     & 2 $\times$     & 3 $\times$     & 1 $\times$     \\
		WebViewLoadDataWithBaseUrl-UnauthorizedFileAccess                          & 95 $\checkmark$                             & 25 $\checkmark$                              & 17 $\times$     & 1 $\checkmark$ & 1 $\checkmark$ & 4 $\times$     & 2 $\checkmark$ & 1 $\checkmark$ & 1 $\checkmark$ & 5 $\checkmark$ & 2 $\checkmark$ \\
		WebViewOverrideUrl-MITM$^{*}$                                              & 94 $\times$                                 & 26 $\times$                                  & 18 $\times$     & 2 $\checkmark$ & 2 $\checkmark$ & 4 $\checkmark$ & 3 $\times$     & 2 $\checkmark$ & 2 $\checkmark$ & 3 $\checkmark$ & 2 $\checkmark$ \\
		WebViewProceed-UnauthorizedAccess                                          & 102 $\times$                                & 27 $\times$                                  & 18 $\times$     & 2 $\times$     & 3 $\times$     & 5 $\times$     & 3 $\times$     & 2 $\times$     & 1 $\times$     & 4 $\times$     & 2 $\times$     \\
		\cmidrule(lr){1-12} Reported Security Issus                                & 5654                                        & 1405                                         & 978             & 89             & 92             & 193            & 116            & 95             & 82             & 157            & 91             \\
		\cmidrule(lr){1-12} Detected APK Vulnerabilities                           & 11/60                                       & 18/60                                        & 8/60            & 40/60          & 40/60          & 42/60          & 43/60          & 47/60          & 47/60          & 47/60          & 47/60          \\
		\cmidrule(lr){1-12} Vulnerability Recall Rate                              & 18.3\%                                      & 30.0\%                                       & 13.3\%          & 66.7\%         & 66.7\%         & 70.0\%         & 71.7\%         & 78.3\%         & 78.3\%         & 78.3\%         & 78.3\%         \\
		\cmidrule(lr){1-12}
	\end{tabular}
	\label{tab:vuln_detect_summary}
\end{table*}

\begin{table*}
	[t]
	\centering
	\scriptsize
	\caption{Vulnerability exploitation results with LLM configurations (M1,M2,M3): Planner, task
	executor, task validator. Model abbreviations: G2.5F=Gemini-2.5-Flash; G2.5P=Gemini-2.5-Pro. Vulnerability
	name suffixes: *=Benchmark dataset speculative vulnerability findings detected by Gemini-2.5-Pro
	aggregation matching benchmark labels; -{Num}=New aggregated speculative vulnerability findings
	(may include false positives). \colorbox{yellow!20}{Yellow highlighted cells}=False positives (FP),
	otherwise true positives (TP). Status symbols: \ding{72}=TP successfully exploited; \ding{75}=FP
	correctly identified; $\otimes$=TP misclassified as FP; $\odot$=FP misclassified as TP; $\times$=Execution
	error; $\bullet$=Max steps reached. Function call: Total(Successful); task execution: Total(Validated).
	Due to space limitations, the table omits 16 out of scope and 2 unable to install speculative vulnerability
	findings.}
	\label{tab:poc_summary}
	\begin{tabular}{llcccccc}
		\addlinespace[-1.5pt] \cmidrule(lr){1-8} \addlinespace[-1.5pt] \multirow{2}{*}{Type}                                                               & \multirow{2}{*}{Vulnerability Name}                          & \multicolumn{3}{c}{(G2.5P, G2.5F, G2.5F)} & \multicolumn{3}{c}{(G2.5P, G2.5P, G2.5P)} \\
		                                                                                                                                                   &                                                              & Status                                    & Function Call                            & Task Execution & Status    & Function Call & Task Execution \\
		\addlinespace[-1.5pt] \cmidrule(lr){1-8} \addlinespace[-1.5pt] \multirow{9}{*}{Crypto}                                                             & BlockCipher-ECB-InformationExposure$^{*}$                    & \ding{72}                                 & 8 (8)                                    & 5 (5)          & \ding{72} & 5 (5)         & 4 (4)          \\
		                                                                                                                                                   & BlockCipher-ECB-InformationExposure-1                        & \ding{72}                                 & 5 (5)                                    & 4 (4)          & \ding{72} & 5 (5)         & 4 (4)          \\
		                                                                                                                                                   & BlockCipher-ECB-InformationExposure-2                        & \ding{72}                                 & 5 (5)                                    & 3 (3)          & \ding{72} & 5 (5)         & 3 (3)          \\
		                                                                                                                                                   & BlockCipher-NonRandomIV-InformationExposure$^{*}$            & \ding{72}                                 & 15 (14)                                  & 5 (5)          & \ding{72} & 14 (14)       & 5 (5)          \\
		                                                                                                                                                   & BlockCipher-NonRandomIV-InformationExposure-1                & \ding{72}                                 & 12 (12)                                  & 4 (4)          & \ding{72} & 11 (11)       & 5 (5)          \\
		                                                                                                                                                   & ConstantKey-ForgeryAttack$^{*}$                              & \ding{72}                                 & 5 (5)                                    & 3 (3)          & \ding{72} & 7 (6)         & 3 (3)          \\
		                                                                                                                                                   & ExposedCredentials-InformationExposure$^{*}$                 & $\times$                                  & --                                       & --             & \ding{72} & 6 (3)         & 2 (2)          \\
		                                                                                                                                                   & PBE-ConstantSalt-InformationExposure$^{*}$                   & \ding{72}                                 & 33 (29)                                  & 5 (5)          & \ding{72} & 14 (14)       & 5 (5)          \\
		                                                                                                                                                   & PBE-ConstantSalt-InformationExposure-1                       & \ding{72}                                 & 6 (5)                                    & 3 (3)          & \ding{72} & 7 (7)         & 3 (3)          \\
		\addlinespace[-1.5pt] \cmidrule(lr){1-8} \addlinespace[-1.5pt] \multirow{18}{*}{ICC}                                                               & DynamicRegBroadcastReceiver-UnrestrictedAccess$^{*}$         & \ding{72}                                 & 3 (3)                                    & 2 (2)          & \ding{72} & 5 (5)         & 4 (4)          \\
		                                                                                                                                                   & EmptyPendingIntent-PrivEscalation$^{*}$                      & $\otimes$                                 & 10 (8)                                   & 3 (3)          & \ding{72} & 10 (10)       & 5 (5)          \\
		                                                                                                                                                   & FragmentInjection-PrivEscalation$^{*}$                       & \ding{72}                                 & 5 (5)                                    & 2 (2)          & \ding{72} & 8 (8)         & 3 (3)          \\
		                                                                                                                                                   & HighPriority-ActivityHijack$^{*}$                            & \ding{72}                                 & 58 (45)                                  & 6 (5)          & \ding{72} & 13 (13)       & 4 (4)          \\
		                                                                                                                                                   & ImplicitPendingIntent-IntentHijack$^{*}$                     & \ding{72}                                 & 13 (11)                                  & 5 (5)          & \ding{72} & 9 (9)         & 3 (3)          \\
		                                                                                                                                                   & ImplicitPendingIntent-IntentHijack-1                         & \ding{72}                                 & 4 (4)                                    & 3 (3)          & \ding{72} & 4 (4)         & 3 (3)          \\
		                                                                                                                                                   & InadequatePathPermission-InformationExposure$^{*}$           & \ding{72}                                 & 12 (12)                                  & 2 (1)          & \ding{72} & 8 (8)         & 4 (4)          \\
		                                                                                                                                                   & IncorrectHandlingImplicitIntent-UnauthorizedAccess$^{*}$     & \ding{72}                                 & 4 (4)                                    & 7 (7)          & \ding{72} & 4 (4)         & 2 (2)          \\
		                                                                                                                                                   & NoValidityCheckOnBroadcastMsg-UnintendedInvocation$^{*}$     & \ding{72}                                 & 5 (5)                                    & 3 (3)          & \ding{72} & 5 (5)         & 3 (3)          \\
		                                                                                                                                                   & OrderedBroadcast-DataInjection-1                             & \ding{72}                                 & 6 (6)                                    & 2 (2)          & \ding{72} & 10 (10)       & 4 (4)          \\
		                                                                                                                                                   & StickyBroadcast-DataInjection-1                              & $\times$                                  & --                                       & --             & $\times$  & --            & --             \\
		                                                                                                                                                   & TaskAffinity-ActivityHijack-1                                & \ding{72}                                 & 6 (6)                                    & 4 (4)          & \ding{72} & 5 (5)         & 4 (4)          \\
		                                                                                                                                                   & TaskAffinity-PhishingAttack-1                                & \ding{72}                                 & 3 (3)                                    & 3 (3)          & \ding{72} & 7 (7)         & 4 (3)          \\
		                                                                                                                                                   & TaskAffinityAndReparenting-PhishingAndDoSAttack-1            & \ding{72}                                 & 5 (5)                                    & 4 (4)          & \ding{72} & 6 (6)         & 4 (4)          \\
		                                                                                                                                                   & UnhandledException-DOS$^{*}$                                 & \ding{72}                                 & 3 (3)                                    & 2 (2)          & \ding{72} & 4 (4)         & 3 (3)          \\
		                                                                                                                                                   & UnprotectedBroadcastRecv-PrivEscalation$^{*}$                & \ding{72}                                 & 1 (1)                                    & 2 (2)          & \ding{72} & 4 (4)         & 3 (3)          \\
		                                                                                                                                                   & UnprotectedBroadcastRecv-PrivEscalation-1                    & $\bullet$                                 & 117 (113)                                & 12 (9)         & \ding{72} & 29 (29)       & 7 (6)          \\
		                                                                                                                                                   & WeakChecksOnDynamicInvocation-DataInjection$^{*}$            & \ding{72}                                 & 23 (23)                                  & 2 (1)          & \ding{72} & 8 (8)         & 5 (5)          \\
		\addlinespace[-1.5pt] \cmidrule(lr){1-8} \addlinespace[-1.5pt] \multirow{4}{*}{Networking}                                                         & CheckValidity-InformationExposure$^{*}$                      & \ding{72}                                 & 3 (3)                                    & 2 (2)          & \ding{72} & 22 (20)       & 6 (6)          \\
		                                                                                                                                                   & \cellcolor{yellow!20}InvalidCertificateAuthority-MITM-1      & \ding{75}                                 & 1 (1)                                    & 1 (1)          & \ding{75} & 1 (1)         & 1 (1)          \\
		                                                                                                                                                   & \cellcolor{yellow!20}OpenSocket-InformationLeak-1            & \ding{75}                                 & 11 (11)                                  & 6 (3)          & \ding{75} & 11 (11)       & 6 (3)          \\
		                                                                                                                                                   & \cellcolor{yellow!20}OpenSocket-InformationLeak-2            & \ding{75}                                 & 7 (7)                                    & 5 (3)          & $\odot$   & 14 (14)       & 6 (6)          \\
		\addlinespace[-1.5pt] \cmidrule(lr){1-8} \addlinespace[-1.5pt] \multirow{3}{*}{NonAPI}                                                             & MergeManifest-UnintendedBehavior$^{*}$                       & \ding{72}                                 & 4 (4)                                    & 2 (2)          & \ding{72} & 6 (6)         & 3 (3)          \\
		                                                                                                                                                   & MergeManifest-UnintendedBehavior-1                           & \ding{72}                                 & 25 (23)                                  & 3 (3)          & \ding{72} & 10 (10)       & 3 (3)          \\
		                                                                                                                                                   & \cellcolor{yellow!20}OutdatedLibrary-DirectoryTraversal-1    & \ding{75}                                 & 38 (38)                                  & 3 (2)          & \ding{75} & 14 (14)       & 3 (2)          \\
		\addlinespace[-1.5pt] \cmidrule(lr){1-8} \addlinespace[-1.5pt] \multirow{2}{*}{Permission}                                                         & \cellcolor{yellow!20}UnnecesaryPerms-PrivEscalation-1        & \ding{75}                                 & 1 (1)                                    & 1 (1)          & \ding{75} & 1 (1)         & 1 (1)          \\
		                                                                                                                                                   & WeakPermission-UnauthorizedAccess$^{*}$                      & \ding{72}                                 & 7 (6)                                    & 2 (2)          & \ding{72} & 3 (3)         & 3 (2)          \\
		\addlinespace[-1.5pt] \cmidrule(lr){1-8} \addlinespace[-1.5pt] \multirow{13}{*}{Storage}                                                           & ExternalStorage-DataInjection$^{*}$                          & $\bullet$                                 & 63 (58)                                  & 6 (6)          & \ding{72} & 15 (14)       & 7 (6)          \\
		                                                                                                                                                   & ExternalStorage-InformationLeak$^{*}$                        & $\bullet$                                 & 17 (16)                                  & 7 (4)          & $\times$  & --            & --             \\
		                                                                                                                                                   & InternalStorage-DirectoryTraversal$^{*}$                     & \ding{72}                                 & 15 (15)                                  & 2 (2)          & \ding{72} & 8 (8)         & 2 (2)          \\
		                                                                                                                                                   & InternalToExternalStorage-InformationLeak$^{*}$              & \ding{72}                                 & 12 (12)                                  & 4 (3)          & \ding{72} & 8 (8)         & 3 (3)          \\
		                                                                                                                                                   & SQLite-execSQL$^{*}$                                         & \ding{72}                                 & 4 (4)                                    & 2 (2)          & \ding{72} & 5 (5)         & 3 (3)          \\
		                                                                                                                                                   & SQLlite-RawQuery-SQLInjection$^{*}$                          & \ding{72}                                 & 15 (14)                                  & 3 (3)          & \ding{72} & 14 (14)       & 5 (3)          \\
		                                                                                                                                                   & SQLlite-RawQuery-SQLInjection-1                              & \ding{72}                                 & 7 (6)                                    & 4 (4)          & \ding{72} & 10 (10)       & 4 (4)          \\
		                                                                                                                                                   & SQLlite-RawQuery-SQLInjection-2                              & \ding{72}                                 & 4 (4)                                    & 3 (3)          & \ding{72} & 5 (5)         & 4 (4)          \\
		                                                                                                                                                   & SQLlite-RawQuery-SQLInjection-3                              & $\bullet$                                 & 90 (74)                                  & 7 (5)          & \ding{72} & 15 (15)       & 4 (3)          \\
		                                                                                                                                                   & SQLlite-SQLInjection$^{*}$                                   & \ding{72}                                 & 7 (7)                                    & 2 (2)          & \ding{72} & 5 (5)         & 3 (3)          \\
		                                                                                                                                                   & \cellcolor{yellow!20}SQLlite-SQLInjection-1                  & \ding{75}                                 & 1 (1)                                    & 1 (1)          & \ding{75} & 1 (1)         & 1 (1)          \\
		                                                                                                                                                   & SQLlite-SQLInjection-2                                       & \ding{72}                                 & 9 (6)                                    & 3 (3)          & \ding{72} & 9 (9)         & 4 (4)          \\
		                                                                                                                                                   & SQLlite-SQLInjection-3                                       & \ding{72}                                 & 4 (4)                                    & 3 (3)          & \ding{72} & 4 (4)         & 3 (3)          \\
		\addlinespace[-1.5pt] \cmidrule(lr){1-8} \addlinespace[-1.5pt] \multirow{7}{*}{System}                                                             & CheckCallingOrSelfPermission-PrivilegeEscalation$^{*}$       & \ding{72}                                 & 5 (4)                                    & 3 (3)          & \ding{72} & 5 (5)         & 2 (2)          \\
		                                                                                                                                                   & ClipboardUse-InformationExposure$^{*}$                       & \ding{72}                                 & 5 (5)                                    & 6 (3)          & \ding{72} & 3 (3)         & 4 (4)          \\
		                                                                                                                                                   & DynamicCodeLoading-CodeInjection$^{*}$                       & \ding{72}                                 & 12 (11)                                  & 4 (4)          & \ding{72} & 11 (11)       & 4 (4)          \\
		                                                                                                                                                   & EnforceCallingOrSelfPermission-PrivilegeEscalation$^{*}$     & \ding{72}                                 & 19 (18)                                  & 4 (3)          & \ding{72} & 10 (10)       & 4 (4)          \\
		                                                                                                                                                   & EnforcePermission-PrivilegeEscalation-1                      & $\bullet$                                 & 110 (106)                                & 7 (5)          & $\bullet$ & 98 (79)       & 6 (8)          \\
		                                                                                                                                                   & \cellcolor{yellow!20}EnforcePermission-PrivilegeEscalation-2 & \ding{75}                                 & 1 (1)                                    & 1 (1)          & \ding{75} & 1 (1)         & 1 (1)          \\
		                                                                                                                                                   & UniqueIDs-IdentityLeak$^{*}$                                 & \ding{72}                                 & 7 (7)                                    & 4 (3)          & \ding{72} & 5 (5)         & 3 (3)          \\
		\addlinespace[-1.5pt] \cmidrule(lr){1-8} \addlinespace[-1.5pt] \multirow{7}{*}{Web}                                                                & HttpConnection-MITM$^{*}$                                    & \ding{72}                                 & 9 (9)                                    & 2 (2)          & $\bullet$ & 58 (56)       & 7 (7)          \\
		                                                                                                                                                   & JavaScriptExecution-CodeInjection$^{*}$                      & \ding{72}                                 & 11 (11)                                  & 4 (4)          & \ding{72} & 10 (10)       & 5 (5)          \\
		                                                                                                                                                   & UnsafeIntentURLImpl-InformationExposure$^{*}$                & $\otimes$                                 & 17 (12)                                  & 6 (5)          & \ding{72} & 7 (7)         & 5 (5)          \\
		                                                                                                                                                   & WebView-NoUserPermission-InformationExposure$^{*}$           & $\bullet$                                 & 73 (67)                                  & 6 (4)          & \ding{72} & 9 (9)         & 5 (5)          \\
		                                                                                                                                                   & WebViewAllowFileAccess-UnauthorizedFileAccess$^{*}$          & \ding{72}                                 & 5 (5)                                    & 2 (2)          & $\bullet$ & 58 (56)       & 7 (5)          \\
		                                                                                                                                                   & WebViewInterceptRequest-MITM-1                               & \ding{72}                                 & 4 (4)                                    & 3 (3)          & \ding{72} & 2 (2)         & 3 (3)          \\
		                                                                                                                                                   & WebViewOverrideUrl-MITM$^{*}$                                & \ding{72}                                 & 12 (12)                                  & 3 (3)          & \ding{72} & 9 (8)         & 3 (3)          \\
		\addlinespace[-1.5pt] \cmidrule(lr){1-8} \addlinespace[-1.5pt] \multicolumn{2}{l}{Statistics (TP / FP / Out of Scope / Unable to Install / Total)} & \multicolumn{6}{c}{56 / 7 / 17 / 2 / 82}                      \\
		\addlinespace[-1.5pt] \cmidrule(lr){1-8} \addlinespace[-1.5pt] \multicolumn{2}{l}{Average Function Calling Per Task}                               & \multicolumn{3}{c}{4.49}                                     & \multicolumn{3}{c}{2.92}                   \\
		\multicolumn{2}{l}{Function Calling Success Rate}                                                                                                  & \multicolumn{3}{c}{927 / 1002 (92.5\%)}                      & \multicolumn{3}{c}{649 / 680 (95.4\%)}     \\
		\multicolumn{2}{l}{Task Execution Pass Rate}                                                                                                       & \multicolumn{3}{c}{195 / 223 (87.4\%)}                       & \multicolumn{3}{c}{222 / 233 (95.3\%)}     \\
		\multicolumn{2}{l}{Validated Vulnerability Finding Rate (TP / (Total - FP))}                                                                       & \multicolumn{3}{c}{46 / 75 (61.3\%)}                         & \multicolumn{3}{c}{51 / 75 (68.0\%)}       \\
		\addlinespace[-1.5pt] \cmidrule(lr){1-8} \addlinespace[-1.5pt]
	\end{tabular}
\end{table*}

\end{document}